\documentclass[printer]{aa}
\usepackage{natbib}
\bibpunct{(}{)}{;}{a}{}{,} 
\usepackage{graphicx}
\usepackage{tabularx}

\newcommand{\masyr}{$\mathrm{mas}\ \mathrm{yr}^{-1}$}
\newcommand{\kms}{\hbox{km~s$^{-1}$}}
\newcommand{\obj}{CFBDSIR~2149 }
\newcommand{\Mjup}{M$_{\rm Jup}$ }
\newcommand{\Rjup}{R$_{\rm Jup}$ }
\newcommand{\Teff}{$T_{\rm eff}$ }

\begin{document}

\title{CFBDSIR~2149-0403: young isolated planetary-mass object or high-metallicity low-mass brown dwarf?\thanks{}} 
 

\author{
  P. Delorme  \inst{1,2} 
  T. Dupuy \inst{3} 
  J. Gagn\'e  \inst{4,5} 
  C. Reyl\'e \inst{6} 
   T. Forveille \inst{1,2}
  Michael C. Liu \inst{7}
  E. Artigau \inst{8} 
  L. Albert  \inst{8}
  X. Delfosse \inst{1,2}
  F. Allard \inst{9}
  D. Homeier \inst{9}
   L. Malo  \inst{8,10}
   C. Morley \inst{11}
 M.E. Naud \inst{8} 
 M.Bonnefoy \inst{1,2} 
}

\offprints{P. Delorme, \email{Philippe.Delorme@obs.ujf-grenoble.fr} 
Based on observations obtained with X-Shooter on
  VLT-UT2 at ESO-Paranal(run 091.D-0723 ). Based on observations obtained with HAWKI on VLT-UT4 (run 089.C-0952, 090.C-0483, 091.C-0543,092.C-0548,293.C-5019(A)
and run 086.C-0655(A)).  Based on observations obtained with ISAAC on
  VLT-UT3 at ESO-Paranal(run 290.C-5083). Based on observation
  obtained with WIRCam at CFHT (program 2012BF12). Based on \textit{Spitzer} Space telescope DDT observation (program 10166).}

\institute{Univ. Grenoble Alpes, IPAG, F-38000 Grenoble, France
 \and CNRS, IPAG, F-38000 Grenoble, France  
  \and The University of Texas at Austin, Department of Astronomy, 2515 Speedway, Stop C1400, Austin, Texas 78712-1205, USA
    \and Carnegie Institution of Washington DTM, 5241 Broad Branch Road NW, Washington, DC 20015, USA
\and NASA Sagan fellow
   \and Institut UTINAM, CNRS UMR 6213, Observatoire des Sciences de l’Univers THETA Franche-Comt\'e Bourgogne, Univ. Bourgogne Franche-Comt\'e, 41 bis avenue de l’Observatoire
 \and IfA, University of Hawai'i, 2680 Woodlawn Drive, Honolulu, HI 96822, USA
\and Institut de Recherche sur les Exoplan\`etes,
  Universit\'e de Montr\'eal, C.P. 6128, Succursale Centre-Ville,
  Montr\'eal, QC H3C 3J7, Canada
  \and Univ Lyon, Ens de Lyon, Univ Lyon1, CNRS, Centre de Recherche Astrophysique de Lyon UMR5574, F-69007, Lyon, France
  Cedex 07, France
   \and Canada-France-Hawaii Telescope Corporation, 65-1238 Mamalahoa 
   Highway, Kamuela, HI96743, USA
 \and  UC Santa Cruz, ISB 159 / 1156 High St, Santa Cruz, CA 95060 
}

\abstract{}
{We conducted a multi-wavelength, multi-instrument observational characterisation of the candidate free-floating planet CFBDSIR~J214947.2$-$040308.9, a late T-dwarf with possible low-gravity features, in order to constrain its physical properties.}
{ We analyzed 9 hours of X-Shooter spectroscopy with signal detectable from 0.8--2.3~$\mu$m, as well as additional photometry in the mid-infrared using the \textit{Spitzer Space Telescope}.  Combined with a VLT/HAWK-I astrometric parallax, this enabled a full characterisation of the absolute flux from the visible to 5~$\mu$m, encompassing more than 90\% of the expected energy emitted by such a cool late T-type object. Our analysis of the spectrum also provided the radial velocity and therefore the determination of its full 3-D kinematics.}
 {While our new spectrum confirms the low gravity and/or high metallicity of CFBDSIR~2149, the parallax and kinematics safely rule out membership to any known young moving group, including AB~Doradus. We use the equivalent width of the \ion{K}{I} doublet at 1.25$\mu$m as a promising tool to discriminate the effects of low-gravity from the effects of high-metallicity on the emission spectra of cool atmospheres. In the case of CFBDSIR~2149, the observed K\,{\sc i} doublet clearly favours the low-gravity solution. }
{CFBDSIR~2149 is therefore a peculiar late-T dwarf that is probably a young, planetary-mass object (2--13~\Mjup, $<$500~Myr) possibly similar to the exoplanet 51~Eri~b, or perhaps a 2--40~\Mjup brown dwarf with super-solar metallicity.}

\date{}


\authorrunning{P. Delorme et al.}
\titlerunning{In-depth physical characterisation of a candidate free-floating planet }
\maketitle

\section{Introduction}

Brown dwarfs and giant exoplanets populate the same temperature range and share many physical properties, such as their molecule-dominated atmospheres and gradual cooling from $\sim$3000~K at formation to $\sim$100~K like the Solar System gas-giant planets. Recent discoveries of very massive planets \citep{Chauvin.2005,Marois.2010,Delorme.2013}, some possibly more massive than the 13~\Mjup\ deuterium burning mass limit, hint that planets could overlap with brown dwarfs in mass. On the other hand, the discovery of isolated L~dwarfs in young clusters \citep{Zapatero.2002,Ramirez.2012,Zapatero.2014}, in young moving groups \citep{Liu.2013,Gagne.2015c,Gauza.2015}, and very cold very nearby Y~dwarf objects \citep[e.g.,][]{kirkpatrick.2012,Luhman.2014} show that very low-mass isolated brown dwarfs exist and overlap with the planetary masses. When these low-mass brown dwarfs are close enough and bright enough to be observed spectroscopically their atmospheres are much easier to study than similar exoplanets that lie near their very bright host stars. \citet{Liu.2013} notably showed that the  $\sim$8~\Mjup\ brown dwarf PSO~J318.5$-$22, a $\beta$-pictoris moving group member shares the spectral characteristics of the young directly imaged exoplanets, as well as atypically red late-L spectral type objects \citep[e.g.,][]{Faherty.2013,Schneider.2014,Gizis.2015,Kellogg.2016,Schneider.2016,Bonnefoy.2016}. When CFBDSIR~J214947.2$-$040308.9 was identified \citep{Delorme.2012a}, hereafter CFBDSIR~2149, it seemed to be a candidate member of the AB~Doradus young moving group and, together with the low gravity features in its spectrum, made it a unique T-type isolated planetary-mass candidate. Another earlier type isolated young planetary-mass T-dwarf, SDSS~J111010.01+011613.1, has been identified as a bona fide member of  AB~Doradus moving group \citep[$149^{+51}_{-19}$~Myr;][]{Bell.2015} by \citet{Gagne.2015c}. The late-T spectral type of CFBDSIR~2149 is typical of the coolest known directly imaged exoplanets, such as GJ~504~b or 51~Eri~b \citep{Kuzuhara.2013,Macintosh.2015}, that the latest generation of adaptive optics systems are detecting. We therefore carried out a multi-wavelength, multi-instrument follow-up of CFBDSIR~2149 to fully characterise it and constrain its nature.\\

In Section~2 we present the new observations of CFBDSIR~2149, and in Section~3 we discuss the possible membership of \obj to young moving groups using updated kinematic data. In Section~4, we analyse the atmospheric properties that are compatible with this new spectral information, and in Section~5 we combine the spectral information, absolute flux measurement and dynamical information to assess several hypotheses on the physical nature of this peculiar late-T object.
 While we explore in the following the possible surface gravities that can be compatible with the observed spectrum of CFBDSIR~2149, we use the adjectives "low", "intermediate" and "high" gravity. For clarity and self-consistency, we systematically use the term "low gravity" to apply to all scenarii that would correspond to \obj firmly belonging to the planetary mass range, "intermediate" to all scenarii that would lead to a mass at the planet/brown dwarf boundary and "high" to all gravity clearly associated to brown dwarf masses. In practical, this means log $g$=3.5 and 4.0] are "low" gravity, log $g$=4.5 is "intermediate" and  log $g$=5.0 and 5.5 are "high" gravity.

\section{New observations  of CFBDSIR~2149 \label{observations}  }
  \subsection{Spectroscopy \label{spectrobs}}
  We measured the flux emitted by CFBDSIR~2149 from 0.6\,$\mu$m to 2.4\,$\mu$m using 9 hours of X-Shooter \citep{Vernet.2011} observations acquired over several weeks in service mode, resulting in an on-source effective integration time of 7 hours. We used a slit 0.9$\arcsec$ wide both in the visible and near-infrared (hereinafter NIR), with a resolution of 8800 and 5300 in the visible and NIR, respectively.  Individual exposures were 670s in the visible and 234s in the NIR. The seeing was better than 1$\arcsec$ and the airmass below 1.3

 The spectra were reduced using the latest ESO X-Shooter
  pipeline \citep{Modigliani.2010}, which produces 2-dimensional,
  curvature-corrected, spectra from X-Shooter's NIR arm (from 0.99--2.5~$\mu$m) and visible arm (from 0.6--1.02~$\mu$m) for each Observing Block (OB). No signal
  was retrieved for wavelengths shorter than $\sim$0.8~$\mu$m, but a low
  signal-to-noise ratio (SNR) spectrum of the optical far-red was
  recovered from 0.8--1.0~$\mu$m. 
 The trace was extracted with custom IDL
  procedures that used Gaussian boxes in the spatial dimension at each point along
  the spectral direction. The noise spectrum was obtained by measuring  the dispersion among 10 spectral pixels on a noise trace obtained by
  subtracting the science trace from itself after a 1-pixel shift. Since
  the shift is much smaller than the full spectral resolution (4.2
  pixels in the NIR and 6.0 pixels in the visible), this effectively removes
  the science spectrum, but keeps the information on the actual
  background and photon noise from the science trace.

  The resulting 1-D spectra from all 9 OBs were then divided by 
  the telluric spectrum obtained from the observation of standard stars just after or just before each
  OB and that were reduced and extracted using the same pipeline as the science
  OBs. We then refined the spectral calibration for each OB by using the known spectral position of telluric features visible in the sky portion of the science data (i.e., the pixels on either side of the target's spectrum). We also applied the correction for the barycentric velocity at this step so that the spectra for all OBs have the same velocity reference before stacking them. Since the data quality of each OB show significant variations related to more or less optimal observing conditions, the individual spectra were finally
  combined weighting by the inverse variance.
The same reduction and extraction procedures were used for the NIR and
visible arms of X-Shooter, but the SNR in the small common
  wavelength interval between the two arms is very low (SNR $\sim$ 1) and highly variable because it covers the transition from the dichroic sending all light to the VIS arm and the  dichroic sending all light to the NIR arm.  Within this  small common wavelength intersection we selected the range where the signal to noise was the better than 1.0 in each arm, and normalised the visible spectra so that the weighted average of the flux in this range is the same in both arms. Since we have no
 $z'$-band photometric detection of CFBDSIR~2149, we cannot calibrate the visible
 spectrum on photometry. We therefore caution that our scaling is not independently calibrated and might be affected by modest systematic errors. The SNR at full resolution (R$\sim$5000) on the $J$-band peak is around 15 per resolution element.


  Though the X-Shooter data reduction pipeline provides a flux-calibrated spectrum, we verified the flux homogeneity of this large wavelength coverage
  spectral data by comparing it with existing WIRCam and NTT
  photometry (see Table~\ref{spec_phot}).  We synthesized the 
  science spectrum colours by integrating it multiplied by the WIRCam global
  transmission, 
  including filter, instrument and telescope transmission and the
  detector quantum efficiency \citep[see section 2.2 of ][for details]{Delorme.2008b}. We anchored these colours to the $J$-band photometry to obtain
  the spectrophotometric magnitudes. This test shows that the synthesized photometry agrees with our measurements within 2$\sigma$, therefore validating the NIR absolute fluxes measured by X-Shooter.
  As shown in Table \ref{spec_phot}, we also derived
  spectrophotometric $CH_{4\rm on}$ and $CH_{4\rm off}$ magnitudes from the
  spectra.

  We used our X-Shooter spectrum to derive the spectral
  indices defined in \citet{Burgasser.2006}, \citet{Warren.2007}, and \citet{Delorme.2008a}
  that trace the strength of several molecular absorption features
  in T dwarfs. As shown in Table \ref{indices}, the
  spectral indices are typical of a T7.5 dwarf, with a
  significantly enhanced $K/J$ index, telltale of relatively weak
 collision induced absorption by H$_2$ \citep[though greenhouse effect could participate
  to $K$-band flux enhancement, see][]{Allard.2012}, and therefore
  implying a low-pressure photosphere
  \citep{Leggett.2002,Burgasser.2004,Golimowski.2004,Knapp.2004,Burgasser.2006}.
\citet{Hiranaka.2012} propose an alternative explanation for the
similarly red spectral energy distribution of some peculiar L-dwarfs,
which could be caused by a thin dust layer above the
photosphere. Since most of the dust is condensated in late-T dwarf
photospheres, this alternative hypothesis is less likely for objects
as cool as CFBDSIR~2149, making a low-pressure photosphere a more probable explanation for its red $J-Ks$ colour.
 Such a low pressure can be the sign of a young, low-mass and therefore low-gravity object and/or of a more opaque, higher altitude photosphere typical of a high-metallicity object. \\
 
\begin{table*}
\caption{Value of the NIR spectral indices of CFBDSIR~2149
  \citep[as defined by][]{Burgasser.2006,Warren.2007,Delorme.2008a} for some known late-T brown dwarfs.\label{indices}}     
\begin{tabular}{|l|c c c c c c c c c|} \hline   
  Object       &Sp. Type&   H$_2$O-J &W$_J$ &      CH$_4$-J &  H$_2$O-H  &      CH$_4$-H   &      NH$_3$-H& 	 CH$_4$-K & K/J\\ \hline \hline 
  {\bf CFBDSIR~2149} &  T7.5  & 0.067 & 0.328 & 0.209         &  0.228          & 0.143  &   0.640         &0.140  & 0.193\\ 
                  &        & $\pm$0.003  & $\pm$0.005 &$\pm$0.004  &$\pm$0.007&$\pm$0.005  & $\pm$0.005  & $\pm$0.018  &  $\pm$0.003  \\ 
                  &        & T7.5 &  T7.5 & T7.5 & T7.5 & T7.5 & -  & T6.5&  - \\ \hline
SDSS1504+10 &  T7  & 0.082  & 0.416  &  0.342  & 0.241  &  0.184  & 0.668 & 0.126 & 0.132 \\  \hline
Gl~570~D &  T7.5  &  0.059 & 0.330  & 0.208   &0.206  & 0.142   & 0.662  & 0.074 & 0.081  \\ \hline
2M0415 & T8 &  0.030 & 0.310  & 0.172   & 0.172  & 0.106   & 0.618  & 0.067 & 0.133  \\ \hline
Ross458C &  T8+  & 0.007  & 0.269  & 0.202   & 0.219 &  0.107  & 0.701 & 0.082 &  0.192   \\  \hline
Wolf940B&T8+  & 0.030&  0.272 & 0.030 & 0.141   &    0.091   &  
0.537& 0.073 &0.111  \\ \hline 
\end{tabular}
\tablefoot{Values were derived using spectra from this article and from \citet{Burgasser.2006,Delorme.2008a,Burningham.2009,Burningham.2010}. }
\end{table*}

\subsubsection{Radial velocity}

The spectral resolution of our X-Shooter spectrum (R=5300) allowed us to measure the radial velocity of CFBDSIR~2149. 
As described above, our X-Shooter data have been calibrated in velocity, with the barycentric correction applied. Any remaining shift in velocity between our observed spectrum and a properly calibrated reference spectrum would therefore be caused by their respective radial velocities.  
In order to use a reference with minimal noise, we used the best fitting BT-Settl model, with [M/H]=0, log $g$=3.5 and T$_{eff}$=700\,K as a reference. We converted the wavelength dimension into velocity and then cross-correlated the full model with our stacked spectrum of CFBDSIR~2149, taking into account the noise of the observations. We found that \obj has a radial velocity of 8$\pm$4~\kms with respect to the barycenter of the solar system. This measured radial velocity is discrepant at more than 3$\sigma$ with the velocity of a bona-fide AB~Doradus moving group member, strongly challenging the association to this moving group that we proposed in past work \citep{Delorme.2012a}. As a sanity check, we derived the radial velocity of the T7.5 Gl~570~D \citep{Burgasser.2000,McLean.2003,McLean.2007} using its NIRSPEC spectrum at R$\sim$2500 and the same approach. We found a radial velocity of 22$\pm$5~\kms, in good agreement with the radial velocity of 26.8$\pm$0.1~\kms derived by \cite{Nidever.2002} for the much brighter primary Gl~570~A.

 \subsection{NIR Photometry}
\begin{figure*}
\begin{tabular}{cc} \\
\includegraphics[width=8cm]{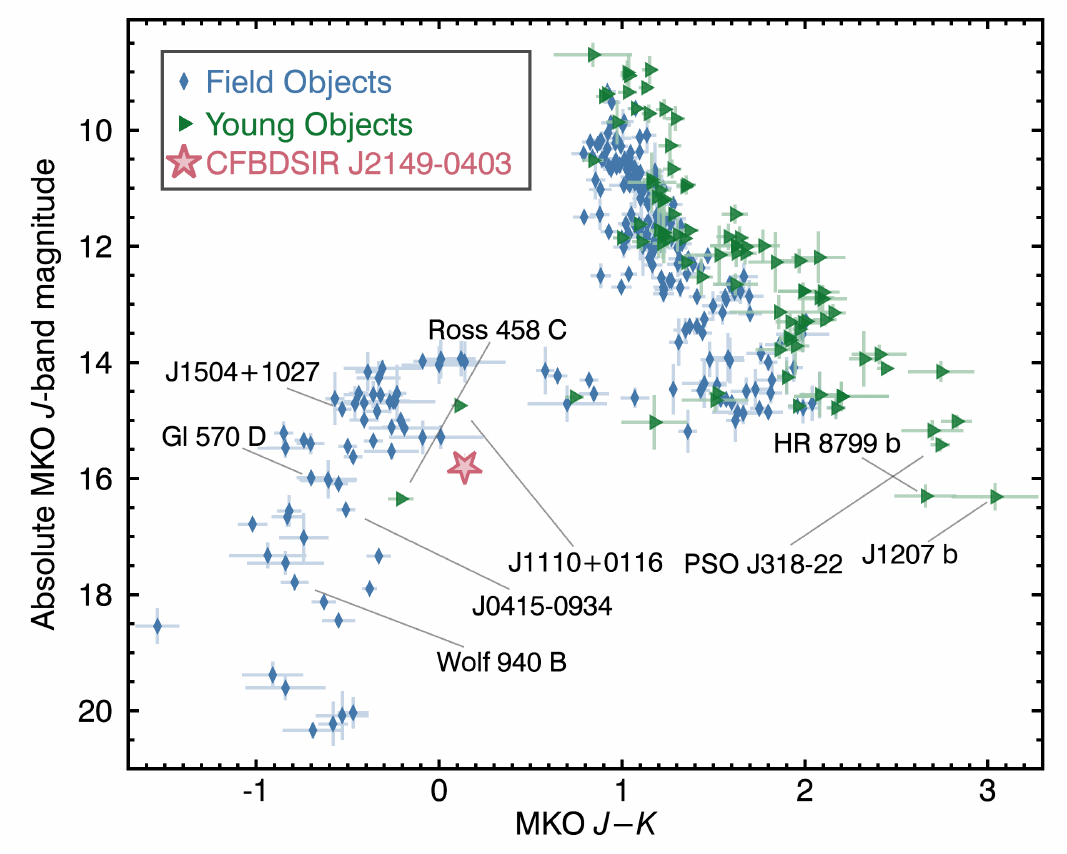} & \includegraphics[width=8cm]{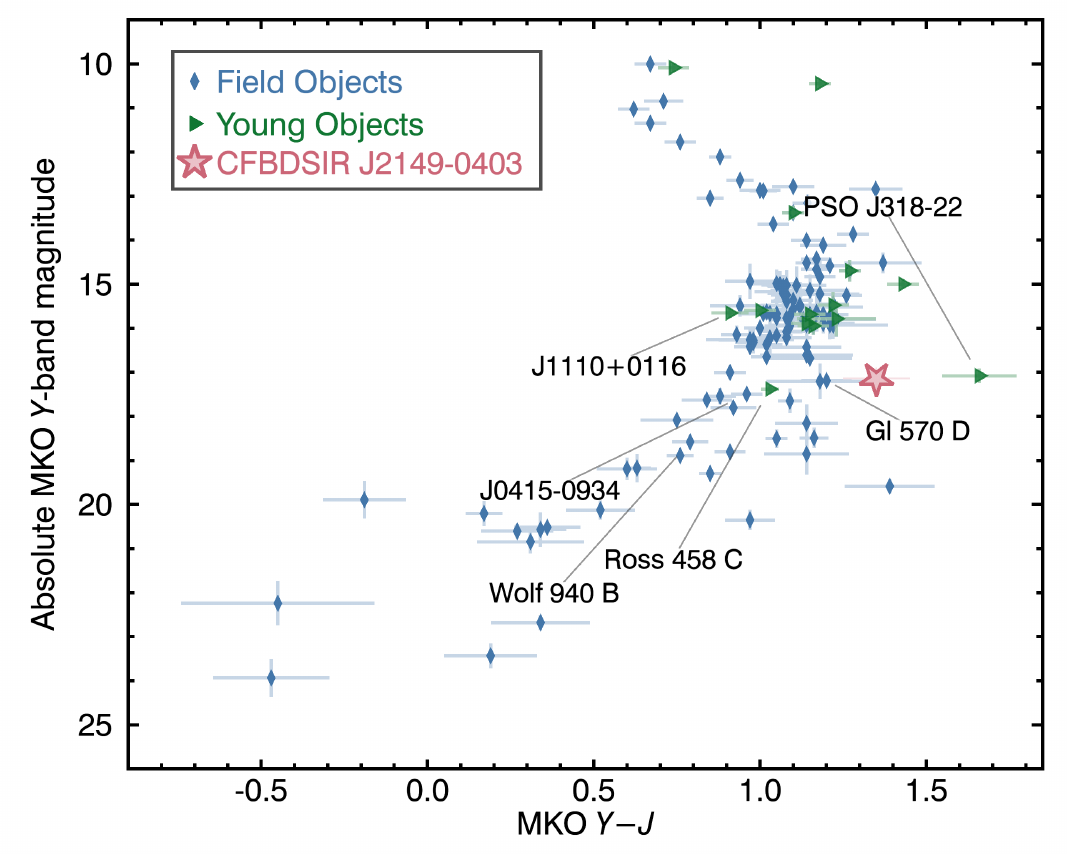} \\
\includegraphics[width=8cm]{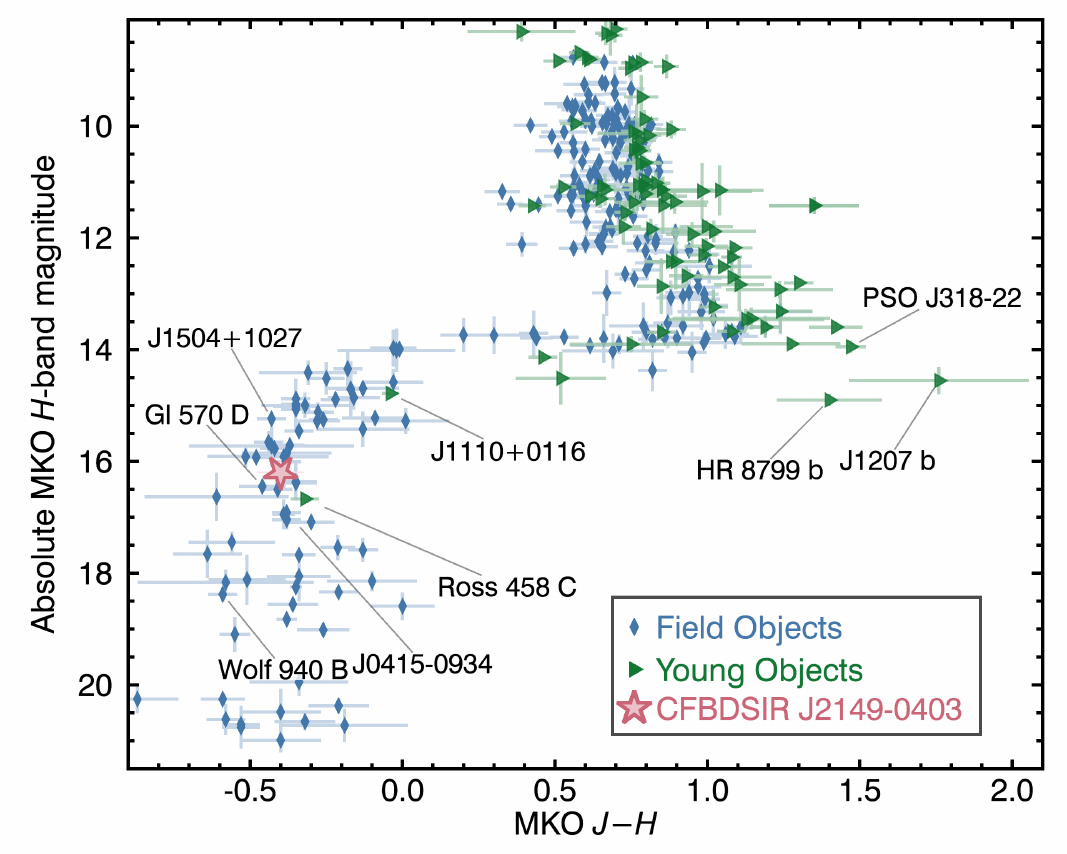} & \includegraphics[width=8cm]{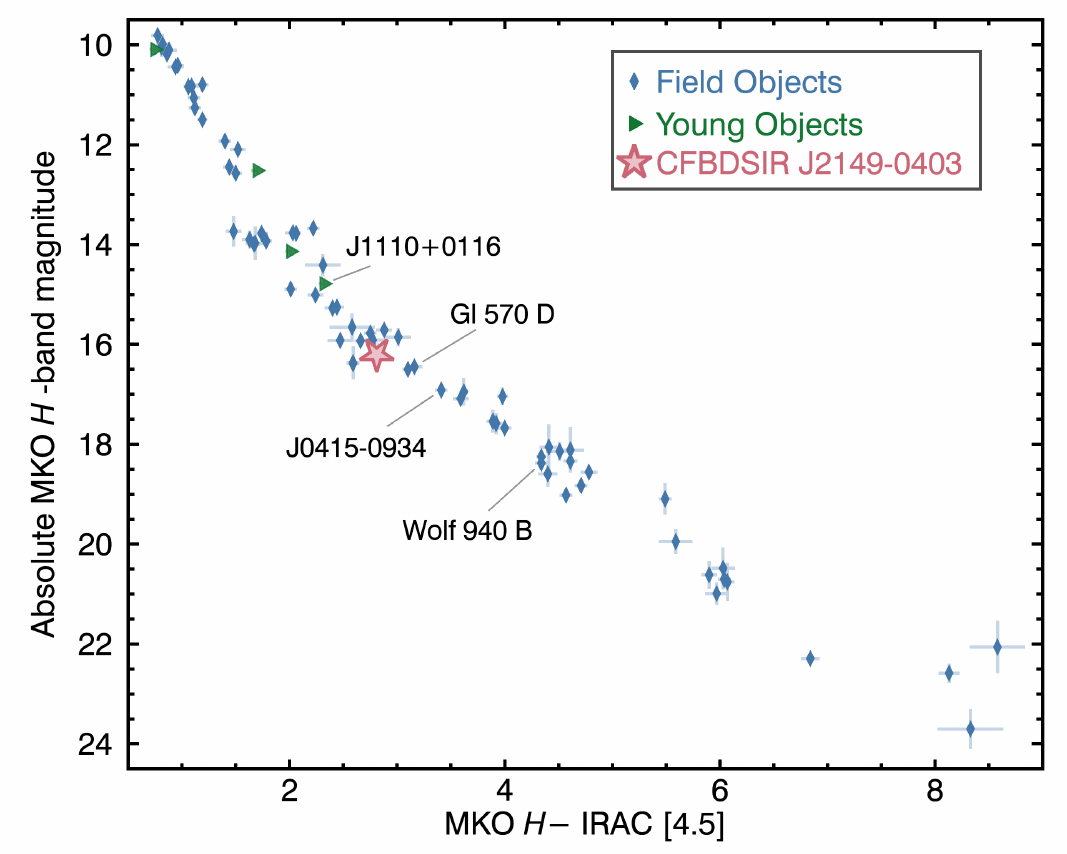} \\
\includegraphics[width=8cm]{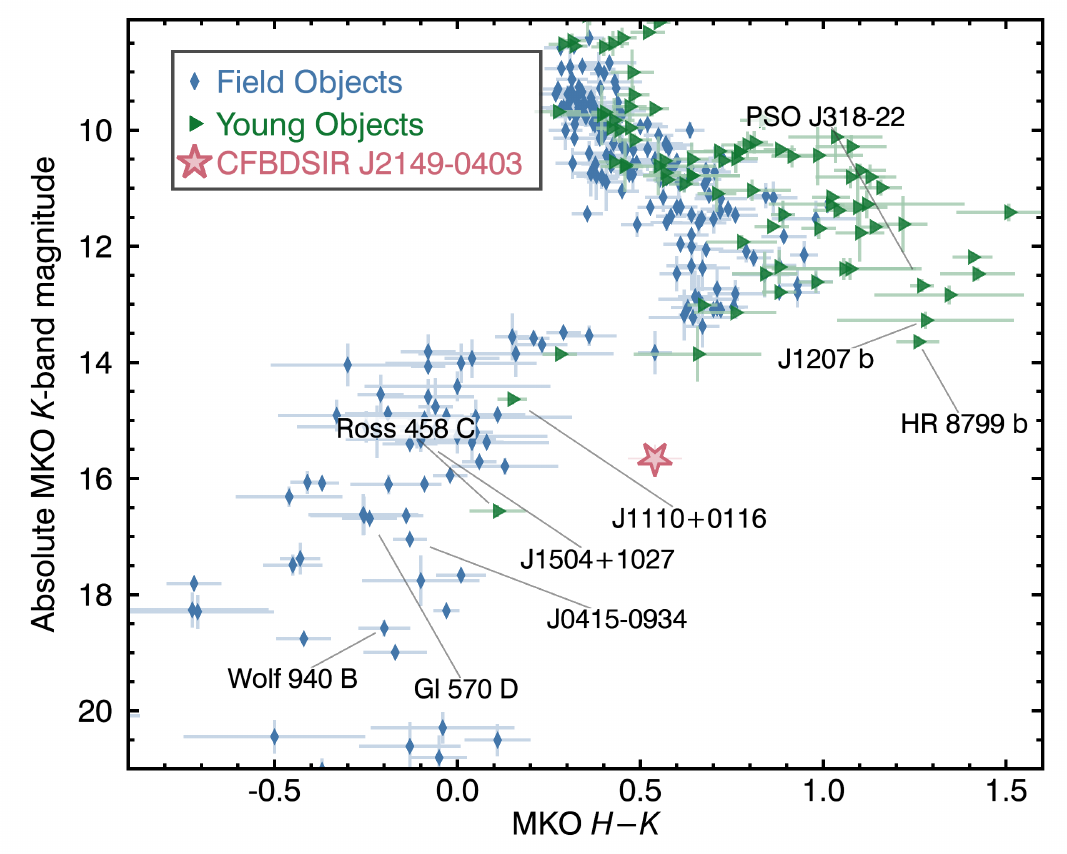} & \includegraphics[width=8cm]{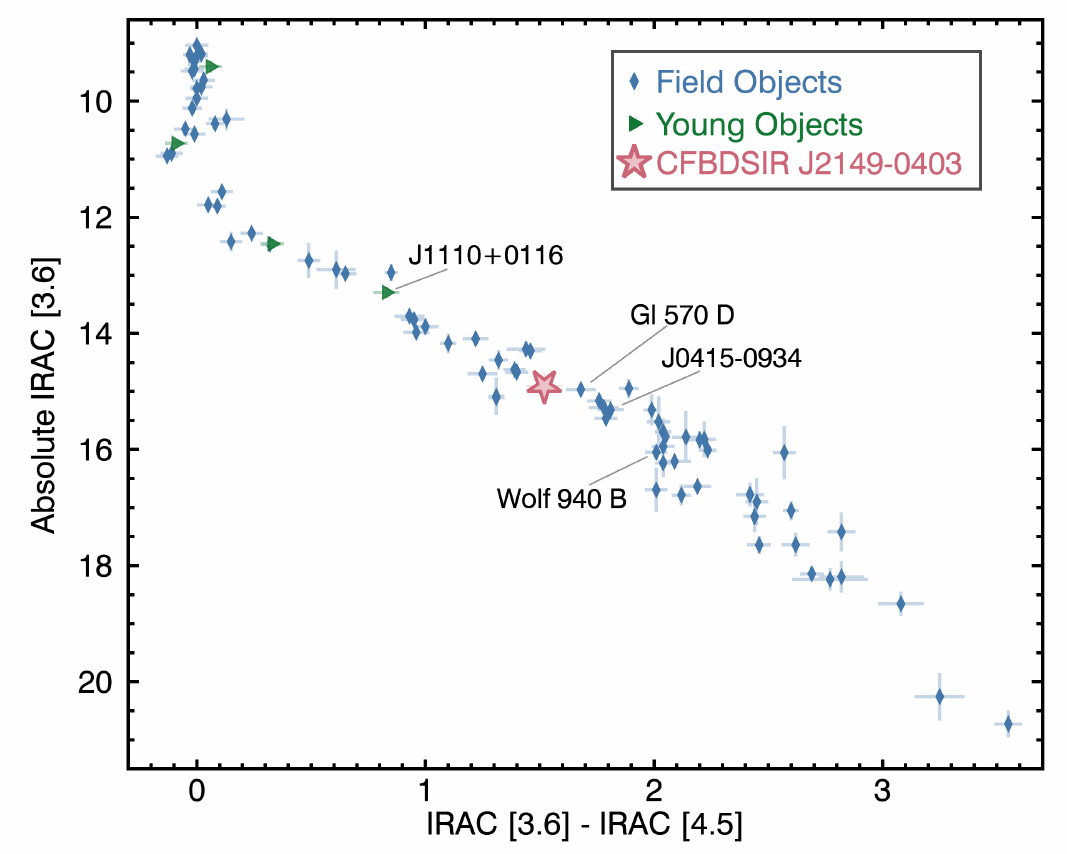} \\
\end{tabular}
 \caption{\label{colspt}Colours and absolutes magnitudes of CFBDSIR~2149 (Red star) compared to
   known field and young L and T dwarfs.}
\end{figure*}

 We also used the wide-field WIRCam imager \citep[20$\arcmin$x20$\arcmin$][]{Puget.2004} at CFHT (run 12BF12) to obtain additional higher SNR photometry in $H$ (800s exposure time, on  2012-09-07, with 0.67$\arcsec$ seeing ) and $Ks$ bands (1360s exposure time, on 2012-09-09 and 2012-10-10 with a seeing of 0.93$\arcsec$ and 0.46$\arcsec$) 
We used a modified version of the \textit{jitter} utility within the ESO Eclipse package \citep{Devillard.2001} to correct for the flat field, subtract the background, and coadd the exposures. We extracted photometry from the resulting images using
point-spread-function fitting within \textit{Source Extractor} \citep{Bertin.1996}, and calibrated the zero point using 2MASS \citep{Skrutskie.2006} stars within the same detector chip as the target, see \citet{Delorme.2012a} for more details. We obtained $Ks=19.34 \pm 0.05$~mag and $ H=19.88 \pm 0.06$~mag, including calibrations errors, therefore confirming the very blue $J-H$ colour of $-0.4$~mag that is typical of late-T dwarfs and the red $J-Ks$ colour of +0.14~mag, very atypical for a late-T dwarf (See Fig \ref{colspt}) and plausibly caused by low gravity or high metallicity.

\subsection{Mid-infrared flux from Spitzer Space Telescope}
Like all late-T dwarfs, \obj is expected to emit most of its  flux  in the thermal infrared, between 3--5~$\mu$m. Obtaining photometric measurments of our targets in the mid-infrared was therefore key to constrain its bolometric luminosity. We initially tried to obtain these crucial photometric constraints in the 3--5~$\mu$m range from the ground (5.5 hours of VLT-ISAAC, run 290.C-5083), but a slightly above average thermal background prevented us from achieving any detection during these deep ground-based observations (after a full reduction and analysis, we only
derived a lower limit on the magnitude, with $L’>$15.7~mag). CFBDSIR~2149 is not detected with WISE and is not in the \textit{Spitzer} archive, so we obtained \textit{Spitzer} observations (Program ID: 10166), in channel 1 (3.6$\mu$m{\bf , hereafter $[3.6]$}) and channel 2 (4.5$\mu$m{\bf , hereafter $[4.5]$}) of IRAC \citep{Fazio.2004}. The data were acquired on 2014-09-02 and the target was clearly detected. We used the basic calibrated data provided by the \textit{Spitzer} archive and MOPEX \citep[MOsaicker and Point source EXtractor][]{Makovoz.2005} to create the mosaic and extract PSF-fitting photometry from our data.
After 500\,s exposure time in each channel, \obj was detected in channel 1 with SNR=14.2, corresponding to a Vega magnitude of $18.59\pm0.07$~mag, and  in channel 2 with SNR=54, corresponding to a Vega magnitude of $17.07\pm0.03$~mag.

\begin{table}
\caption{Photometry and  spectrophotometry of CFBDSIR~2149 (using
  WIRCam/MegaCam filter set to generate synthetic colours). $z'$ is in the
  AB system and all others are in the Vega system. \label{spec_phot}}
\begin{tabular}{|l|cc|} \hline 
Filter & Photometry  &Spectrophotometry$^1$\\ \hline \hline
$z'_{ab}$ $^2$   &   $>23.2$ & 23.96$\pm$0.07   \\
$Y$ $^2$  &  20.83$\pm$0.09$^*$ &  20.88$\pm$0.05  \\
 $J$ & 19.48$\pm$0.04$^*$  &  Reference   \\
  $H $ &   19.88$\pm$0.06 & 19.76$\pm$0.04    \\
  $K_S $&   19.34$\pm$0.05 &  19.44$\pm$0.05  \\
 $CH_{4\rm off}$&   -    &   19.15$\pm$0.04  \\
 $CH_{4\rm on}$ &  20.7$\pm$0.25$^*$  & 20.57$\pm$0.05   \\
 $[3.6]$ &  18.59$\pm$0.07 &   -   \\
 $[4.5]$ &  17.07$\pm$0.03 &   -    \\ \hline
\end{tabular}
\tablefoot{$^1$Spectrophotometry is anchored on
  $J$=19.48$\pm$0.04~mag from WIRCam photometric measurements. 
$^2$~Spectrophotometry is possibly affected by small systematical uncertainties because of data rescaling below 1$~\mu$m.
$^*$ From \citet{Delorme.2012a}  }
\end{table}

\section{ Parallax and Kinematic analysis: does CFBDSIR~2149 belong to the young
  moving group AB Doradus?}
\subsection{Parallax and proper motion}

We monitored \obj with the VLT facility near-infrared imager HAWK-I
\citep{2004SPIE.5492.1763P,2006SPIE.6269E..0WC,2008A&A...491..941K} in
order to measure its parallax and proper motion.  We obtained 9
dithered images in $J$ band at each of 11 epochs over 2.05\,yr
beginning on 2012~Jun~26~UT.  We centered the target on one detector of HAWK-I  (field of view of 3.6$\arcmin$x3.6$\arcmin$ and a pixel scale of 0.106$\arcsec$). We reduced the images using the esorex pipeline (v1.8.13) using calibration data provided by the VLT archive
to subtract darks, divide by a flat, and perform an iterative sky
subtraction masking detected sources.  Astrometric analysis of our
images was performed in a similar manner as described in
\citet{Dupuy.2012} and \citet{Dupuy.2013}.  To
correct non-linear distortion in the astrometry we interpolated the
look-up table provided in the HAWK-I pipeline user manual version 1.9
(VLT-MAN-ESO-19500-4407).  The astrometric reference grid was defined
by 46 other stars in the field-of-view of the detector that \obj was
centered on, 39 of which were in SDSS-DR9 \citep{2012ApJS..203...21A}
and were used for the absolute calibration of the linear terms.  The
FWHM of the target was $0\farcs68\pm0\farcs28$ (median and rms) with
${\rm S/N} = 20$--50 during our observations, which were constrained
to be within 0.05~airmass of transit via timing constraints in the
queue scheduling. 
Since the reference stars in the field have non-zero parallax and proper motion, we used the Besan\c{c}on Galaxy model \citep{Robin.2003} to correct from this small effect (0.26\,mas in parallax and -3\,mas/yr in proper motion) and obtain absolute measurements.

\begin{figure}
\includegraphics[width=9cm]{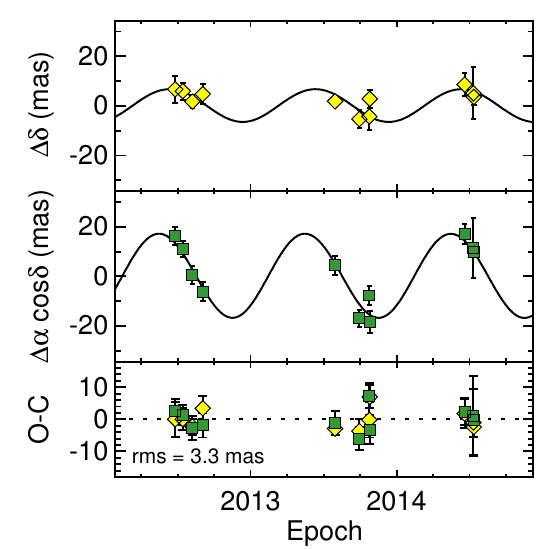} 
 \caption{The top and middle panels show
relative astrometry in Dec ($\delta$) and RA ($\alpha$), respectively,
as a function of Julian year after subtracting the best-fit proper
motion.  (This is for display purposes only; we fit for both the
proper motion and parallax simultaneously in our analysis.)  The
bottom panels show the residuals after subtracting both the parallax
and proper motion.\label{astrom}}.
\end{figure}

 These absolute parallax and proper motion determined by MCMC
are shown in Table~\ref{PM_para} and Figure~\ref{astrom}, and the best-fit solution had a reduced $\chi^2=1.08$.\\

  We also used the 10 good quality astrometric epochs obtained under very good and homegenous conditions to investigate the possible variability of CFBDSIR~2149 in $J$ band. We selected a sample of 100 relatively bright and nearby stars  located in the same HAWK-I detector as our target that we used as flux references. The resulting photometric error was 0.018 magnitude, including the calibration errors caused by the flux dispersion of the reference stars over all epochs, but dominated by error on the point source flux measurement of CFBDSIR~2149 itself. The measured photometric dispersion of \obj over all epochs was 0.022 magnitude, therefore not significantly greater than the expected dispersion arising from photometric and calibration error alone. This  rules out any strong photometric variability (i.e., above about 5\%) in $J$ band for our target over the 10 epochs sampled by our parallax follow-up. However this constraint is too weak to rule out most of the photometric variations currently observed on brown dwarfs, which are usually of lower amplitude \citep[e.g.,][]{Radigan.2014}.  We also note that the new photometric points in $H$ and $Ks$ are within 1$\sigma$ with respect to the photometry reported by \citet{Delorme.2012a}.

\begin{table}
\caption{Proper motion and parallax of CFBDSIR~2149. \label{PM_para}}
\begin{tabular}{|l|c c c|} \hline 
  Parallax &  1-$\sigma$ distance   &  $\mu_{\alpha}$  &  $\mu_{\delta}$ \\
  (mas)  &  range (pc)  & (\masyr)  & (\masyr) \\ \hline \hline 
 18.3$\pm$1.8 & 49.8-60.6  & +138.3$\pm$1.2  & -93.6$\pm$1.5  \\ \hline 
\end{tabular}
\end{table}

\subsection{Young moving group membership  probability }
We used the Bayesian Analysis for Nearby Young AssociatioNs~II tool \citep[BANYAN~II][]{Gagne.2014a} to update the probability that \obj belongs to any young moving group in the solar neighborhood. We used its sky position, proper motion ($\mu_\alpha\cos\delta = 138.3 \pm 1.2$\,\masyr; $\mu_\delta = -93.6 \pm 1.5$\,\masyr), radial velocity ($8 \pm 4$\,\kms) and trigonometric distance ($54.6 \pm 5.4$\,pc) as inputs to BANYAN~II.

The new measurement of proper motion alone had the effect of favoring a membership to the $\beta$~Pictoris moving group ($\beta$PMG) instead of the AB~Doradus moving group (ABDMG), which was reported as most probable by \citet{Delorme.2012a} using the best measurements available then. However, both the trigonometric distance and radial velocity measurements that we present here reject a possible membership to all moving groups considered by BANYAN~II (i.e., ABDMG, $\beta$PMG, Tucana-Horologium, Argus, Columba, Carina and TW~Hydrae). The statistical distances and radial velocities associated with a membership to $\beta$PMG are $24.1 \pm 2.0$\,pc and $-8.2 \pm 1.4$\,\kms, and those to ABDMG are $38.6^{+2.0}_{-2.4}$\,pc and $-11.3 \pm 1.8$\,\kms, when using the updated proper motion measurement and treating radial velocity and distance as marginalized parameters \citep[see][for a detailed explanation on the treatment of marginalized parameters]{Gagne.2014a}.

\begin{figure*}
\begin{tabular}{cc} \\
\includegraphics[width=9cm]{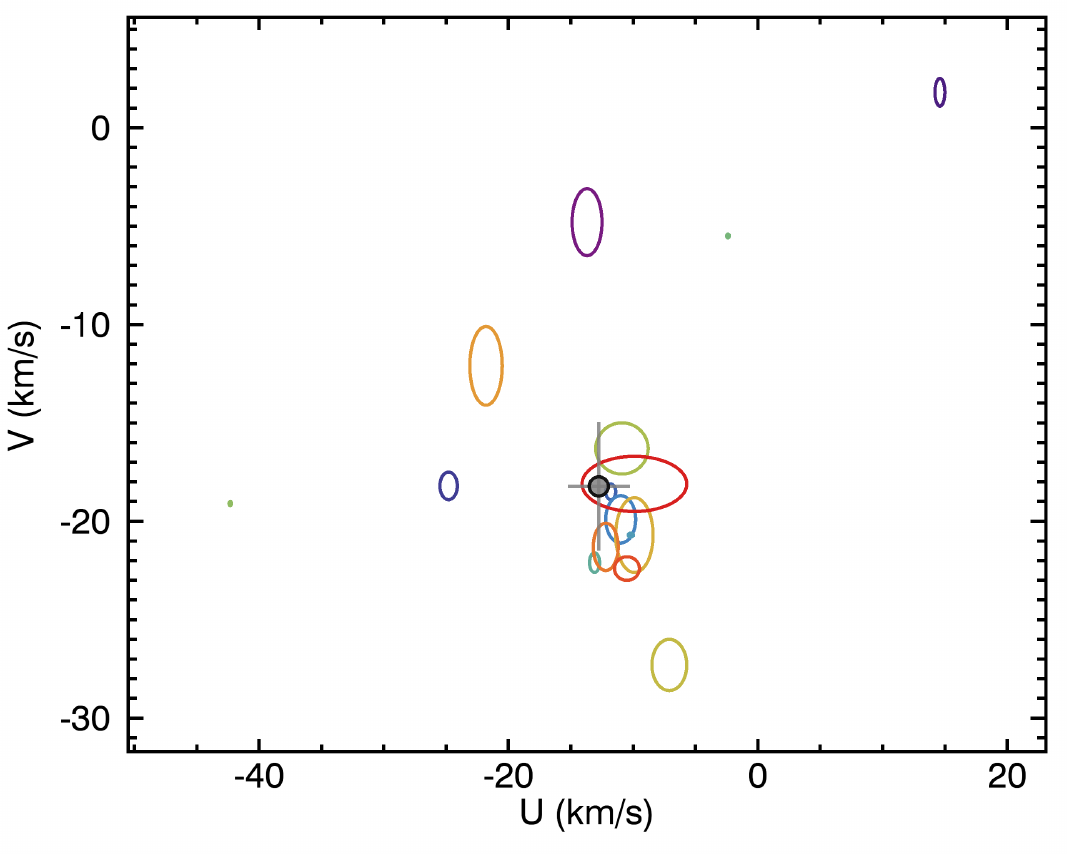} & \includegraphics[width=9cm]{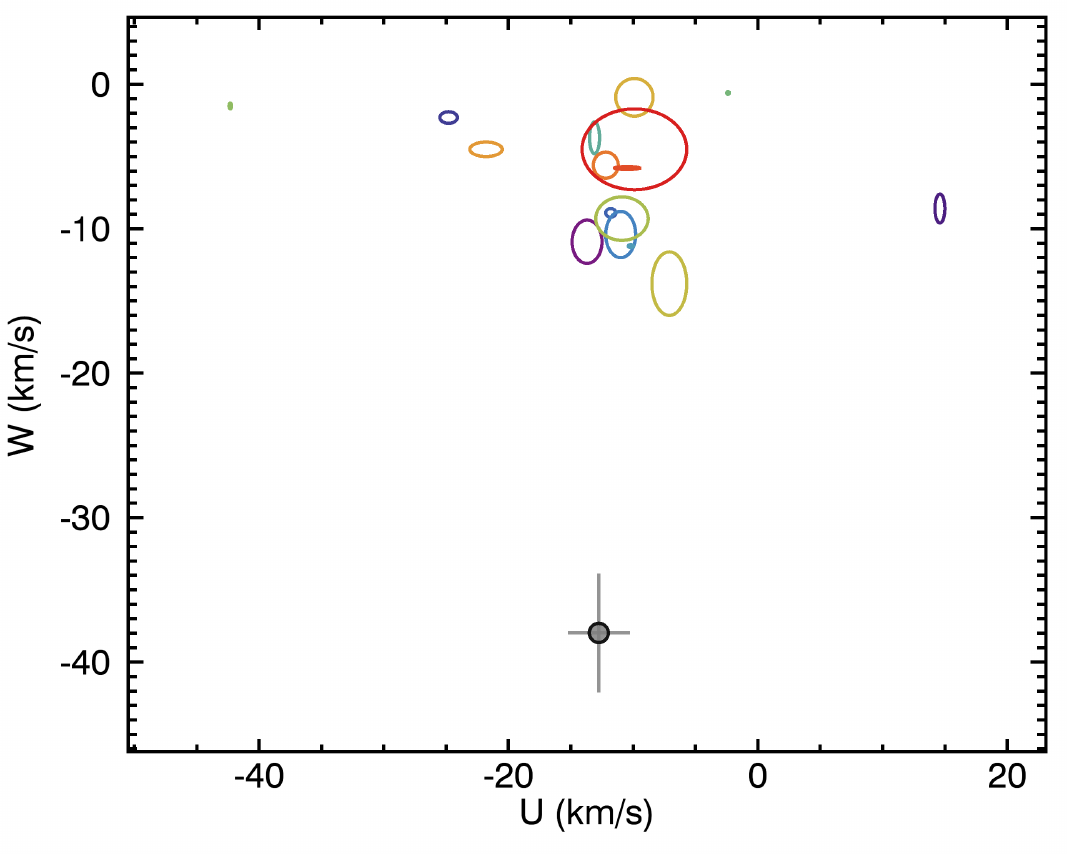} \\
\includegraphics[width=9cm]{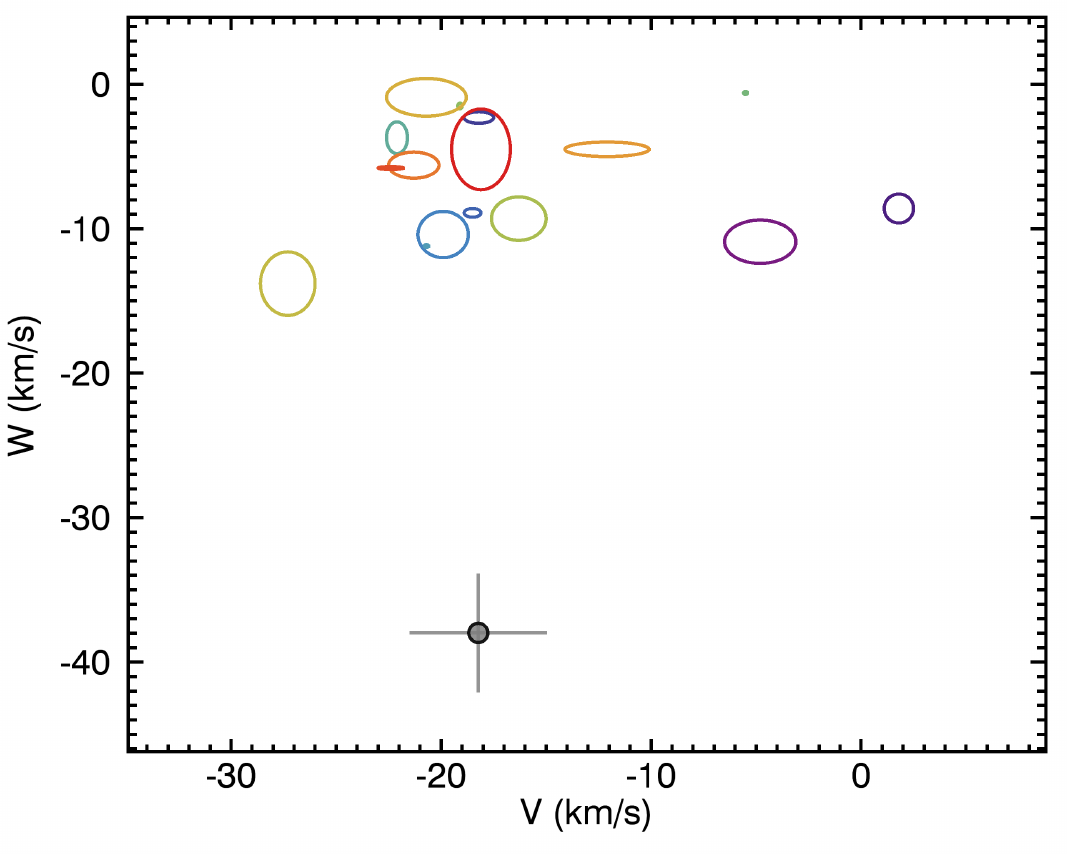}   & \includegraphics[width=9cm]{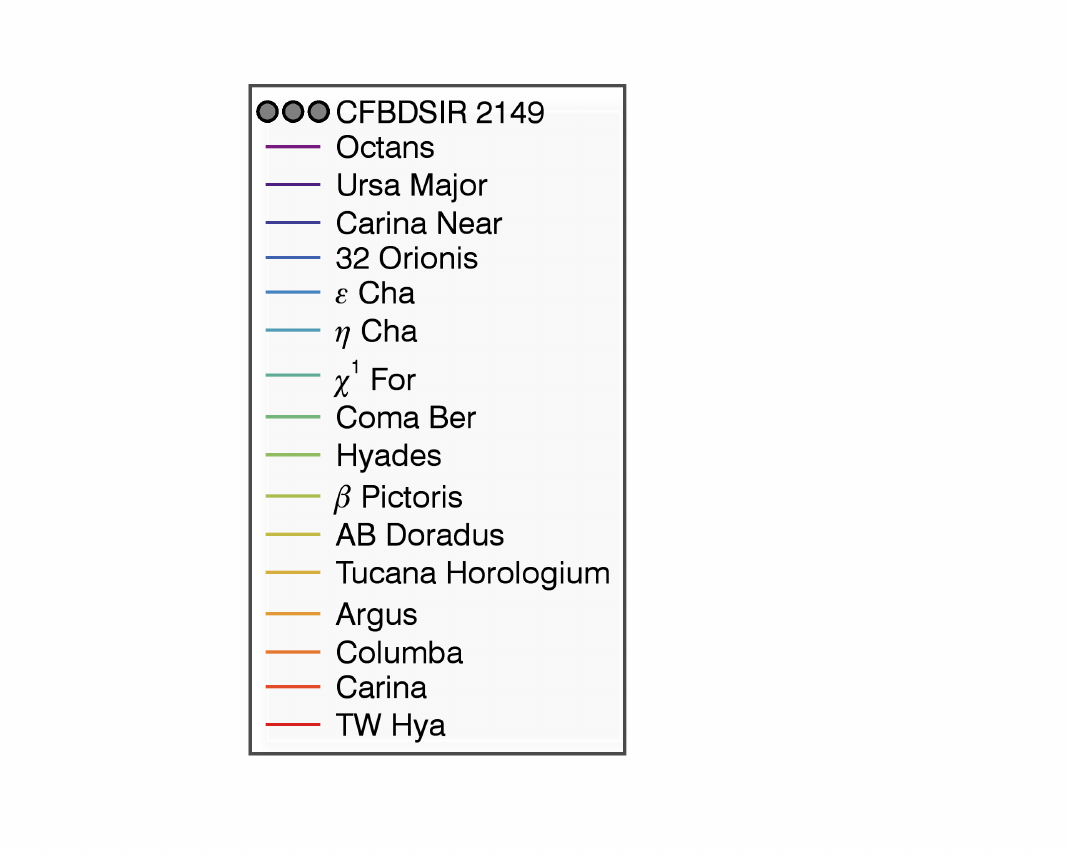}\\
\end{tabular}
 \caption{U, V and W velocities of \obj compared with the 1-$\sigma$ ellipsoid velocity dispersion of the young moving groups considered in BANYAN~II analysis. \label{UVWplot}}.
\end{figure*}

Since we have obtained all measurements needed to compute the $UVW$ space velocity ($-12.8 \pm 2.4$; $-18.2 \pm 3.2$; $-38.0 \pm 4.0$), of CFBDSIR~2149, we can compare it directly with the position and kinematics of more young associations of stars not included in BANYAN~II.  As for \citet{Gagne.2014a}, we use the formalism of \citet{Johnson.1987}, with U positive toward the galactic center, V positive toward the galactic rotation direction and W pointing upward from the galactic plane. The distance between \obj and the distribution mean of these various associations in spatial and kinematic spaces are presented in Fig.~\ref{UVWplot}. We have included similar measurements for the moving groups considered in BANYAN~II for comparison.

No known associations are located within 26\,\kms of CFBDSIR~2149, which corresponds to a minimal 3D Euclidian distance normalized by the $UVW$ scatter of 2.6$\sigma$, see Fig. \ref{UVWplot}. In particular, it can be noted that the $W$ component ($-38.0 \pm 4.0$\,\kms) of the space velocity of \obj taken alone is inconsistent with any known young association with a difference of at least 24\,\kms.  This means that we have no robust dynamical age constraint that we could use as input of the study of the spectra and the atmosphere the object carried out in the following section. All of these considerations hold under the assumption that the young association member distributions follow normal distributions along each of the $XYZUVW$ axes, i.e., that there is no correlation between any combination of $XYZUVW$. For this reason, we used the BANYAN~II moving groups central positions and scatters reported by \citet{Malo.2013}, since those were obtained using the same hypothesis (in contrast with those reported by \citet{Gagne.2014a} that allow for correlations between combinations of $XYZ$ or $UVW$ in the form of rotated Gaussian ellipsoid distributions).

\section{Atmospheric properties of CFBDSIR~2149: a spectral analysis}
\subsection{Spectral Synthesis \label{secspectro}}
   As a first step of the analysis of the atmospheric properties of CFBDSIR~2149, we performed a fit of our full X-Shooter spectrum  beyond 0.8~$\mu$m and IRAC photometry against a grid of BT-Settl 2014 models \citep{Allard.2014proc}, examining effective temperatures from 500\,K to 950\,K with steps of 50\,K and gravities from log $g$ =3.5~dex to 5.5~dex (cgs) with steps of 0.5~dex. We included both solar and super-solar metallicity models (+0.3~dex, i.e., roughly twice the solar abundances). We carried out a simple noise-weighted $\chi^2$ minimisation fit on the full wavelength range, only excluding the range between 1.55--1.59~$\mu$m where BT-Settl models lack the complete methane line list information that has a significant impact on the emerging spectrum of late-T dwarfs \citep[e.g.,][]{Canty.2015}. We also tried limiting the fit to the high SNR emission peaks of our observed spectrum, but it did not affect the final results of the fit, therefore validating our noise-weighted fitting approach.  However we do remove from the fit all data points with very low signal to noise (lower than 0.5) because they could include systematics, and contain negligible information. We first fitted the observed spectra to the models after normalising both observation and models to {\obj the observed $J$-band peak, between 1.255--1.29~$\mu$m}, which is a common approach in brown dwarf studies where parallaxes and hence absolute fluxes are not always available. We then fitted models and observations in absolute fluxes, which is much more constraining on models because they have not only to match the shape of the observed spectrum, but also its actual flux.  \\

\subsubsection{Fitting models to observations after flux normalisation on the $J$-band peak}
\begin{figure*}
\includegraphics[width=18cm]{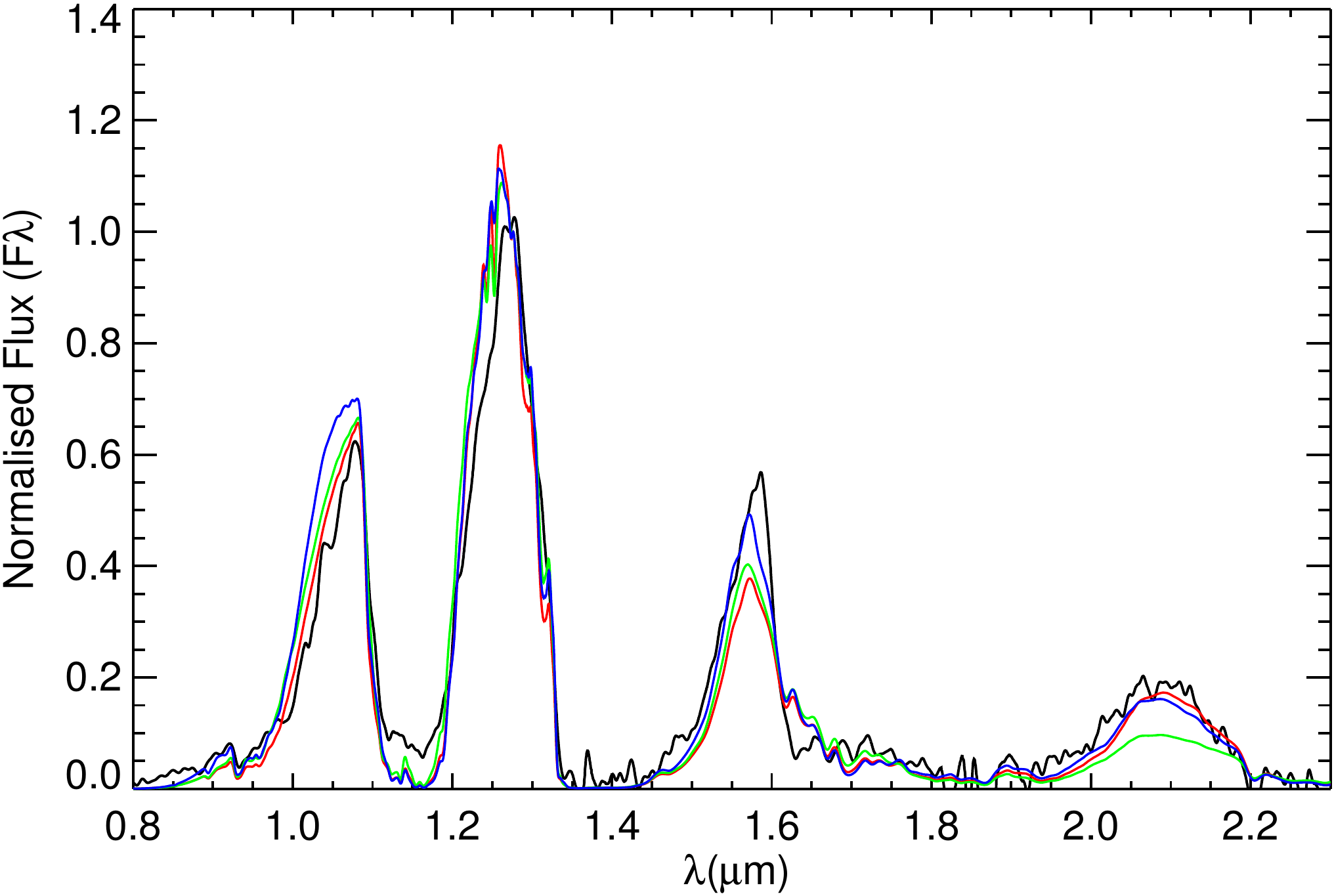}
 \caption{\label{fullrange} Observed spectrum of CFBDSIR~2149 {\bf in black}. Best fitting  BT-Settl models after normalisation on the $J$-band peak flux are represented in colour. The spectra presented here have been binned to R$\sim$200 and are normalised on their $J$-band peak intensity.
{\bf Blue:} Best fitting overall model (700\,K, log $g$=3.5, [M/H]=0).
{\bf Red:} Best fitting  model when high metallicity is forced (800\,K, log $g$=4.5, [M/H]=0.3).
{\bf Green:} Best fitting  model when high gravity and solar metallicity are forced (850\,K, log $g$=5.0, [M/H]=0.0).}
\end{figure*}

 The best fitting model normalised at $J$ band was consistent with our ``by-eye'' fit and corresponded to a cool (700\,K), low gravity (log $g$=3.5) atmosphere at solar metallicity, as shown in blue in Fig. \ref{fullrange}.  The biggest discrepancies between this best fitting model and the observations are in the flux intensity in $Y$-band and in a slight shift in the position of the $J$-band peak, aside from missing CH$_4$ absorption lines in the red part of the $H$-band peak due to an incomplete  CH$_4$ line list in BT-Settl. We also tried to force high metallicity ([M/H]=+0.3) and high gravity (log $g$ $\geq$ 5.0) fits, whose best solutions are also shown in Fig. \ref{fullrange} and Fig.\ref{bandbyband}. The band by band higher resolution spectra of CFBDSIR~2149 and of the best fitting models are shown on Fig.\ref{bandbyband}. While the high gravity solution shows significant discrepancies with respect to the observations in several bands, the high-metallicity solution has almost the same $\chi ^2$ as the the low-gravity one and provides a much better fit in the $Y$ band and a slightly better one in the $K$ band. However the high-metallicity fit also favours moderate gravity (log $g$ =4.5) and leads to a higher effective temperature (800\,K), resulting in underestimating the  strength of H$_2$O and CH$_4$ absorption in $H$ band. The reduced $\chi ^2$ for the best fits are 1.72, 1.73 and 2.06 for the low-gravity, high-metallicity and field-gravity respectively.

\subsubsection{Fitting models to observations in absolute flux}
After obtaining a reliable parallax for CFBDSIR~2149, we were able to directly compare its absolute flux at 54.6$\pm$5.4\,pc to that of models atmospheres that would be located at the same distance. Past studies of brown dwarf atmospheres have often shown that model fitting the shape of an observed spectrum sometimes falls orders of magnitude short of fitting its absolute fluxes, especially when looking at peculiar objects \citep[e.g.,][]{Skemer.2011}.Since the BT-Settl model grid we used is built on the result of the \citet{Baraffe.2003} evolutionary models, each spectra has a physically self-consistent effective gravity adapted to its mass and radius. These physical parameters are shown for the best fitting models on Table \ref{Teffage}. In our case, the best fitting model in absolute flux ($\chi ^2$=2.13) is a 650\,K, low gravity (log $g$=3.5) atmosphere with supersolar metallicity  ([M/H]=+0.3~dex).

We made a 2-D cubic interpolation of the $\chi^2$ values of the relatively coarse models grid (steps of 50~K in effective temperature and 0.5~dex in log $g$) to provide a finer visualisation of the best fitting areas of the parameter range, both for solar metallicity and super solar metallicity models, see Fig.\ref{chimap}.  After this interpolation, the best overall $\chi^2$ (with a value of 1.97) would be for a 680~K, low gravity (log $g$=3.5) solar metallicity object, quite close to the 700~K best fit after normalisation of the flux in the $J$ band. This suggest a slightly lower effective temperature, at solar metallicity, might provide an even better fit to the data than the super solar metallicity solution, with the corresponding best fitting gravity unchanged to log $g$=3.5~dex.

\begin{figure*}
\includegraphics[width=18cm]{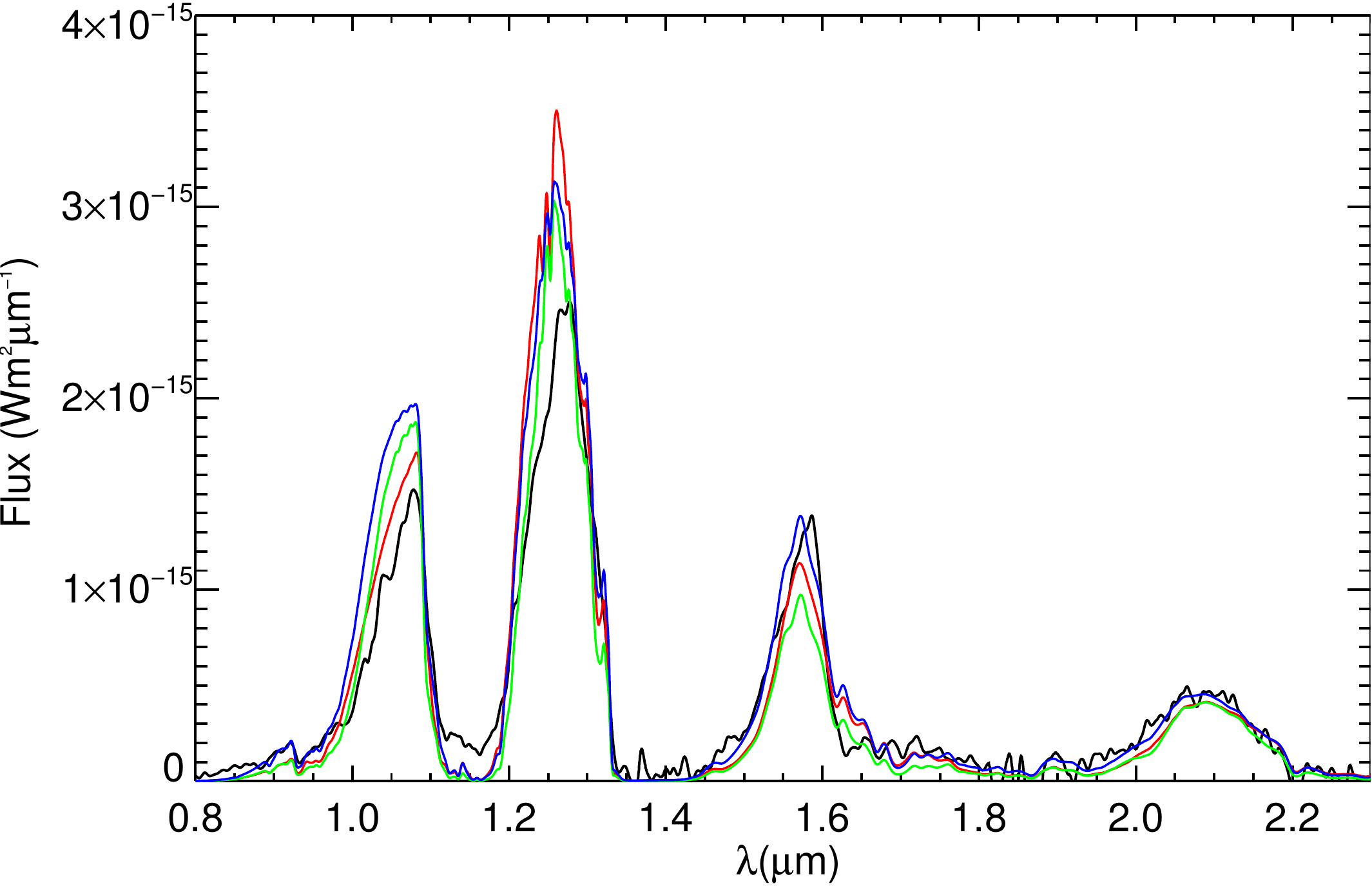}
\caption{\label{fullrange_absolute} Observed spectrum of CFBDSIR~2149 {\bf in black} in each spectral band. Best fitting BT-Settl models in absolute flux ($W.m^2.\mu m^{-1}$) are represented in colour. The spectra presented here have been binned to R$\sim$200.
{\bf Blue:} Model closest to the best fitting solution in absolute flux  after interpolation on both $chi ^2$ maps (700\,K, log $g$=3.5, [M/H]=0).
{\bf Red:} Model closest to the best fitting solution in absolute flux after interpolation on the high-metallicity $chi ^2$ map (800\,K, log $g$=5.0, [M/H]=0.3). 
{\bf Green:} Best fitting model in absolute flux without interpolation (650\,K, log $g$=3.5, [M/H]=0.3).}
\end{figure*}

\begin{figure*}

\begin{tabular}{cc} \\
\includegraphics[width=9cm]{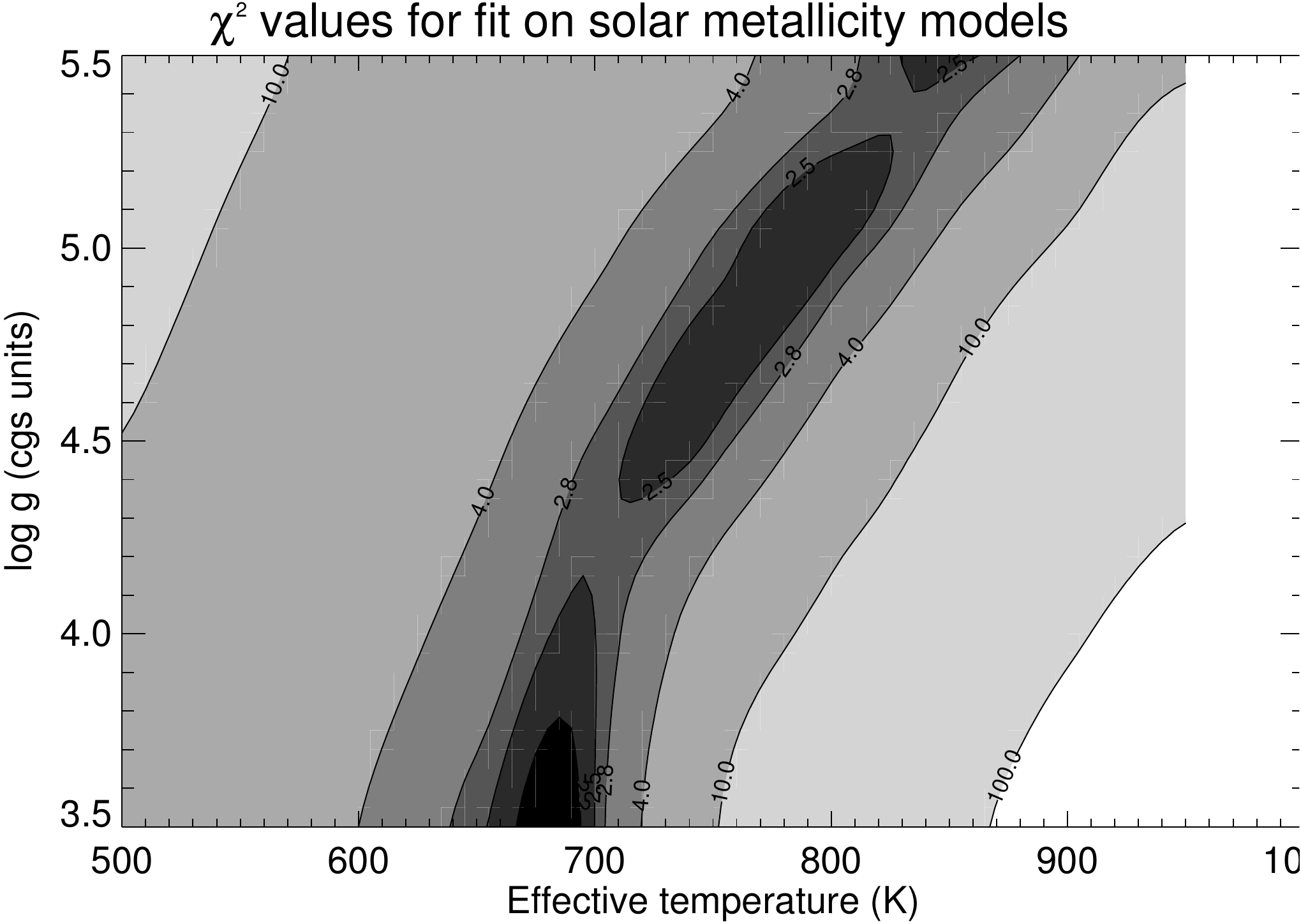}&\includegraphics[width=9cm]{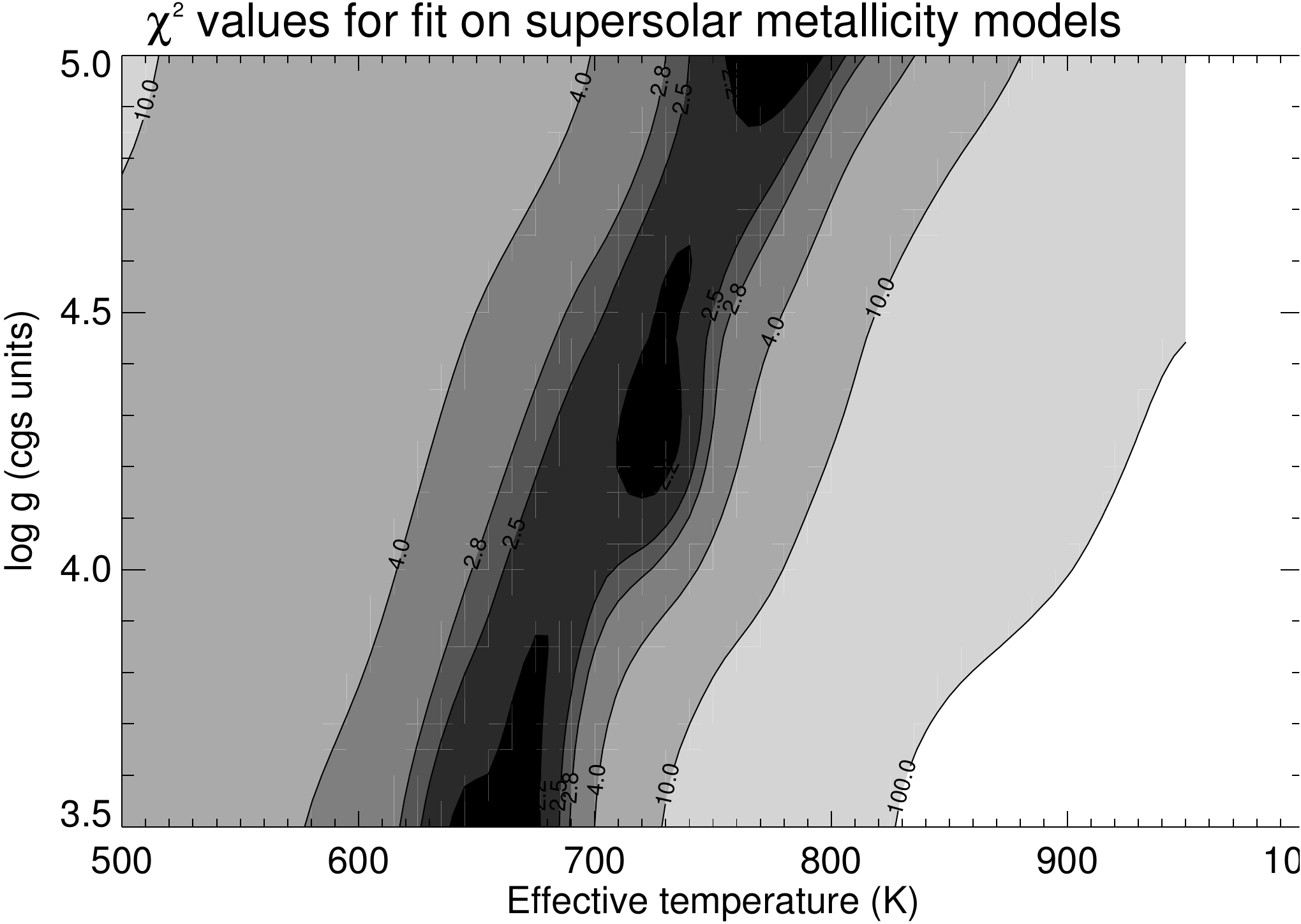}
\end{tabular}
 \caption{\label{chimap} Reduced-$\chi^2$ contour maps for the fit of the observed spectra in absolute flux to the BT-Settl solar (left) and supersolar (right, [M/H]=+0.3) metallicity models. $\chi ^2$ values between the model grid points have been interpolated. The black areas are the best fitting range of the parameter space, where $\chi ^2<$2.2. }
\end{figure*}

\begin{figure*}
\begin{tabular}{cc} \\
\includegraphics[width=9cm]{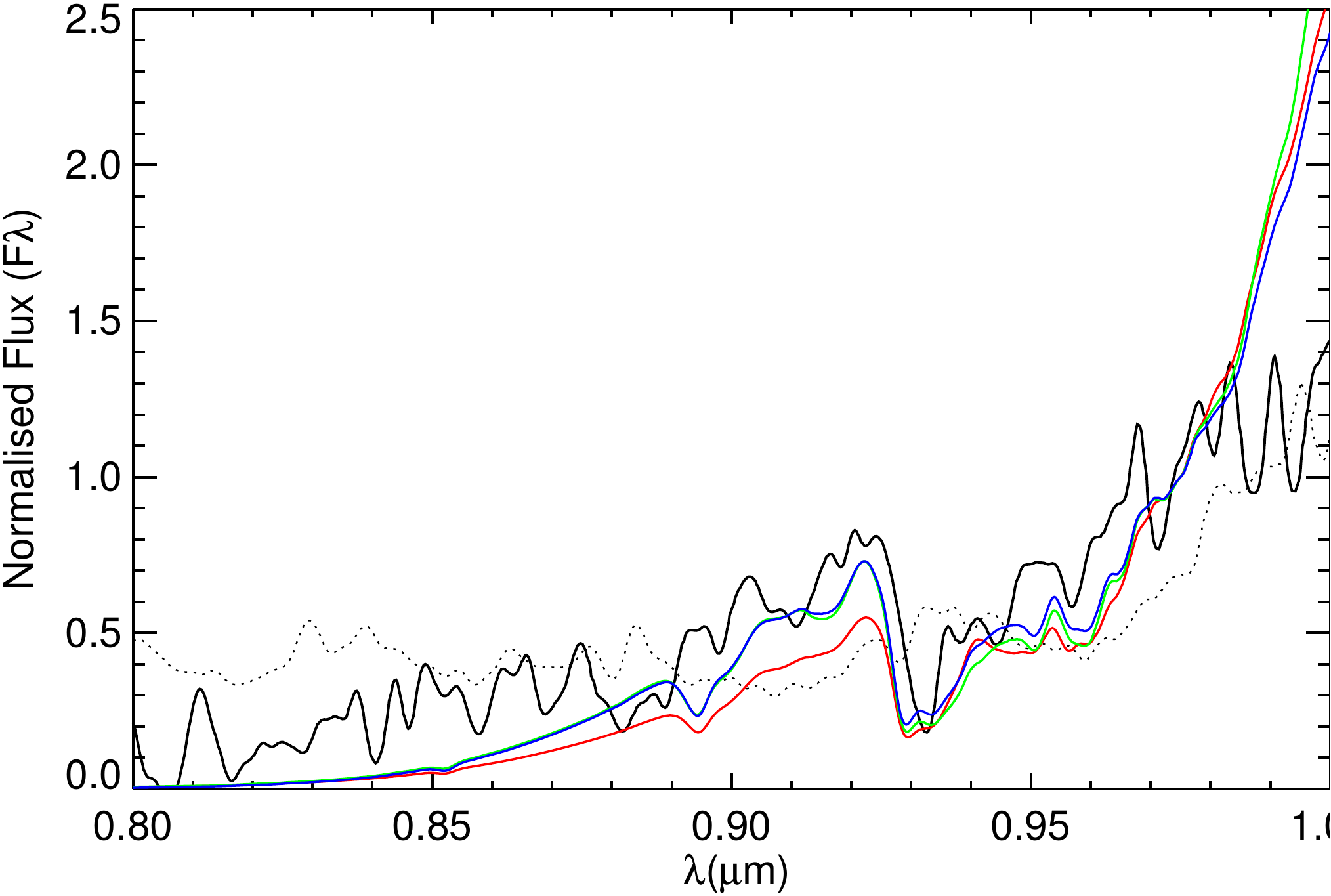} & \includegraphics[width=9cm]{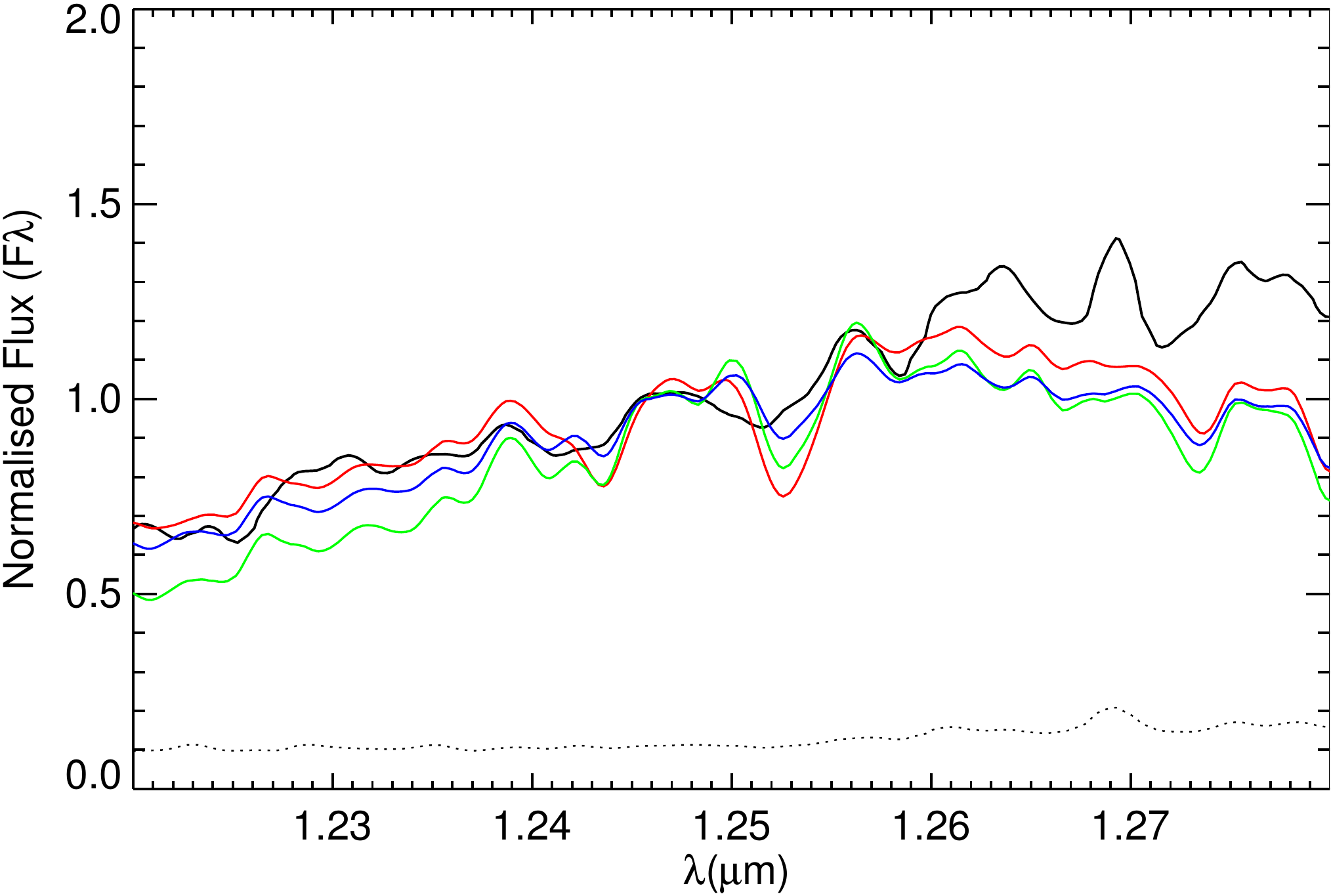} \\
\includegraphics[width=9cm]{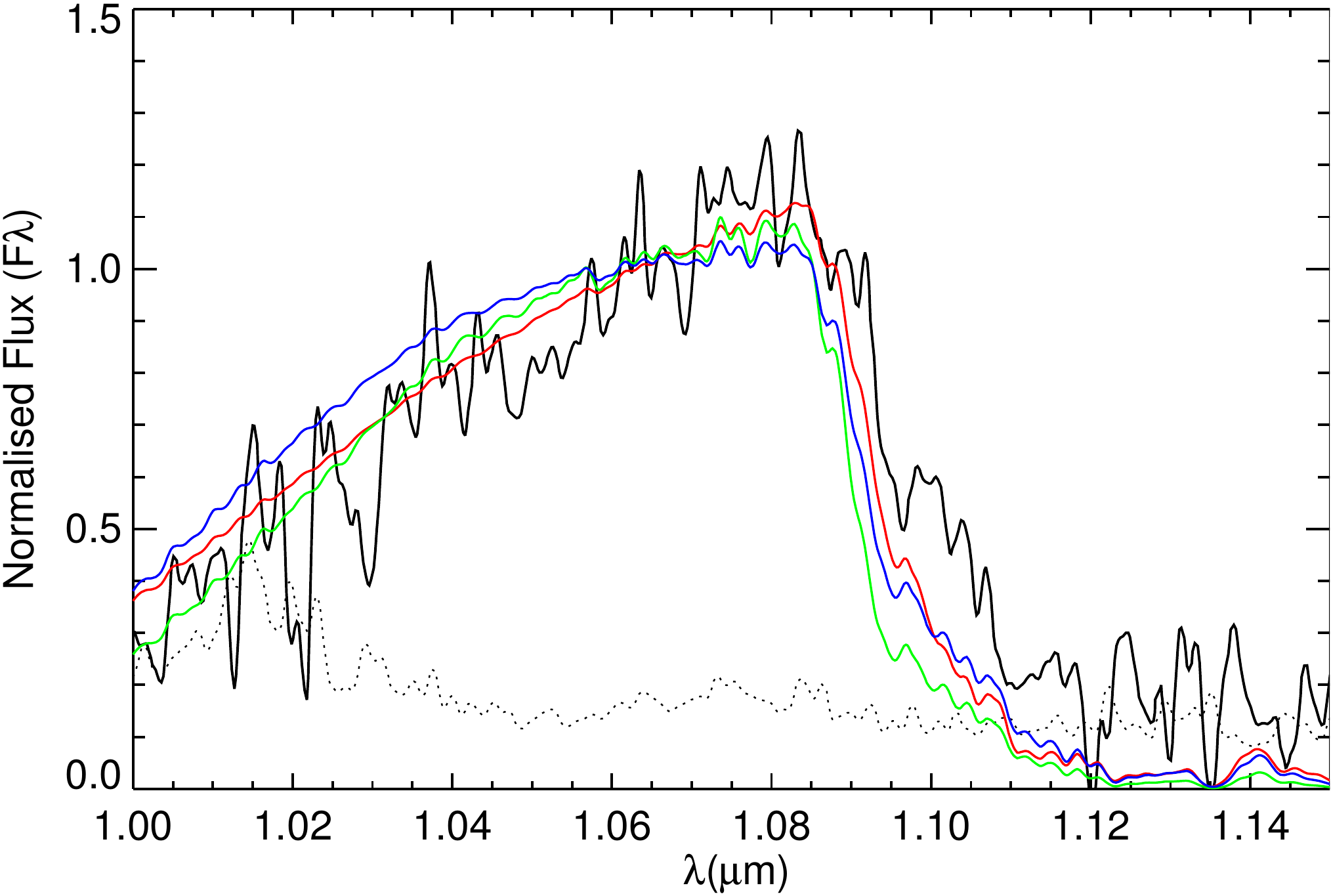}   & \includegraphics[width=9cm]{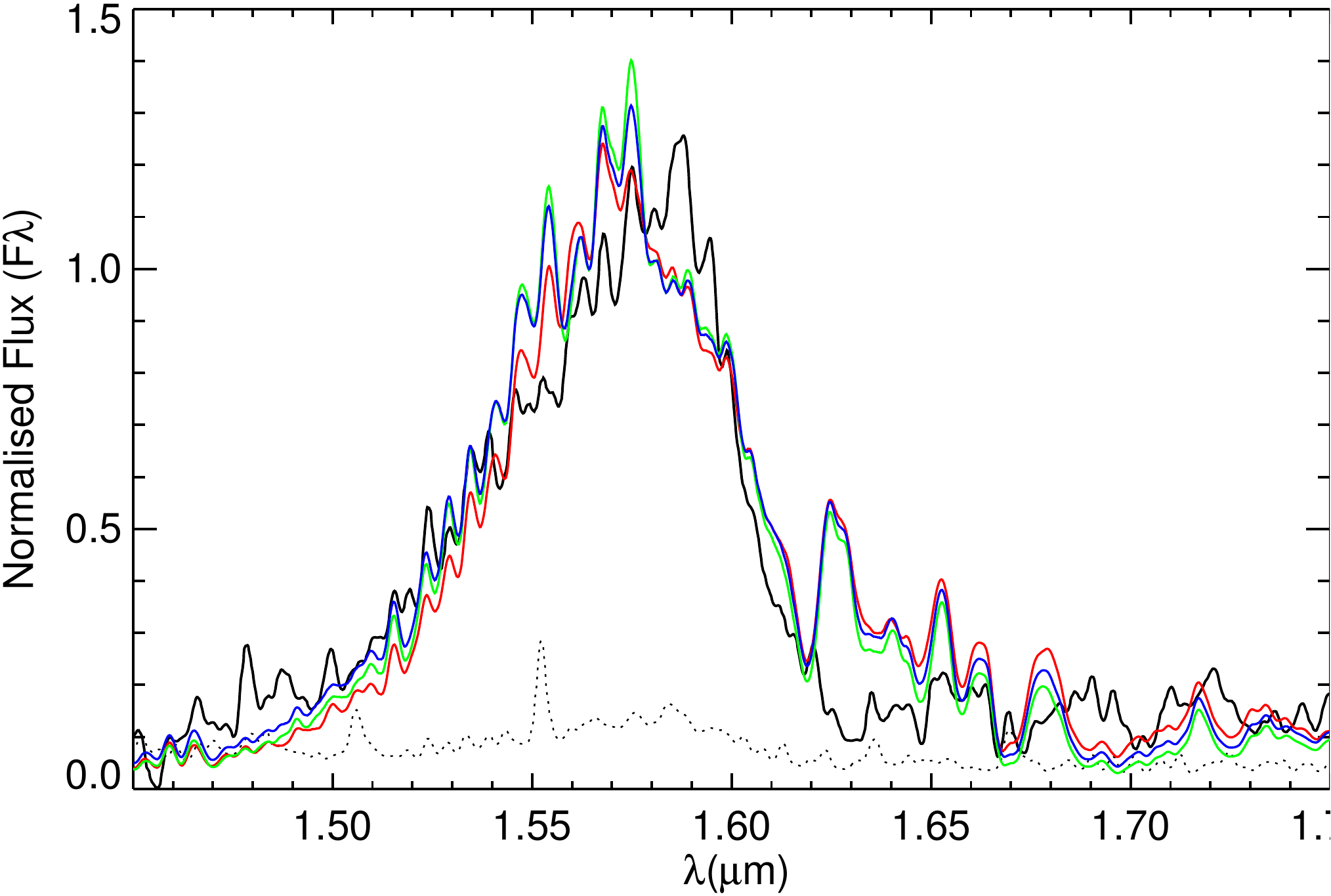}\\
\includegraphics[width=9cm]{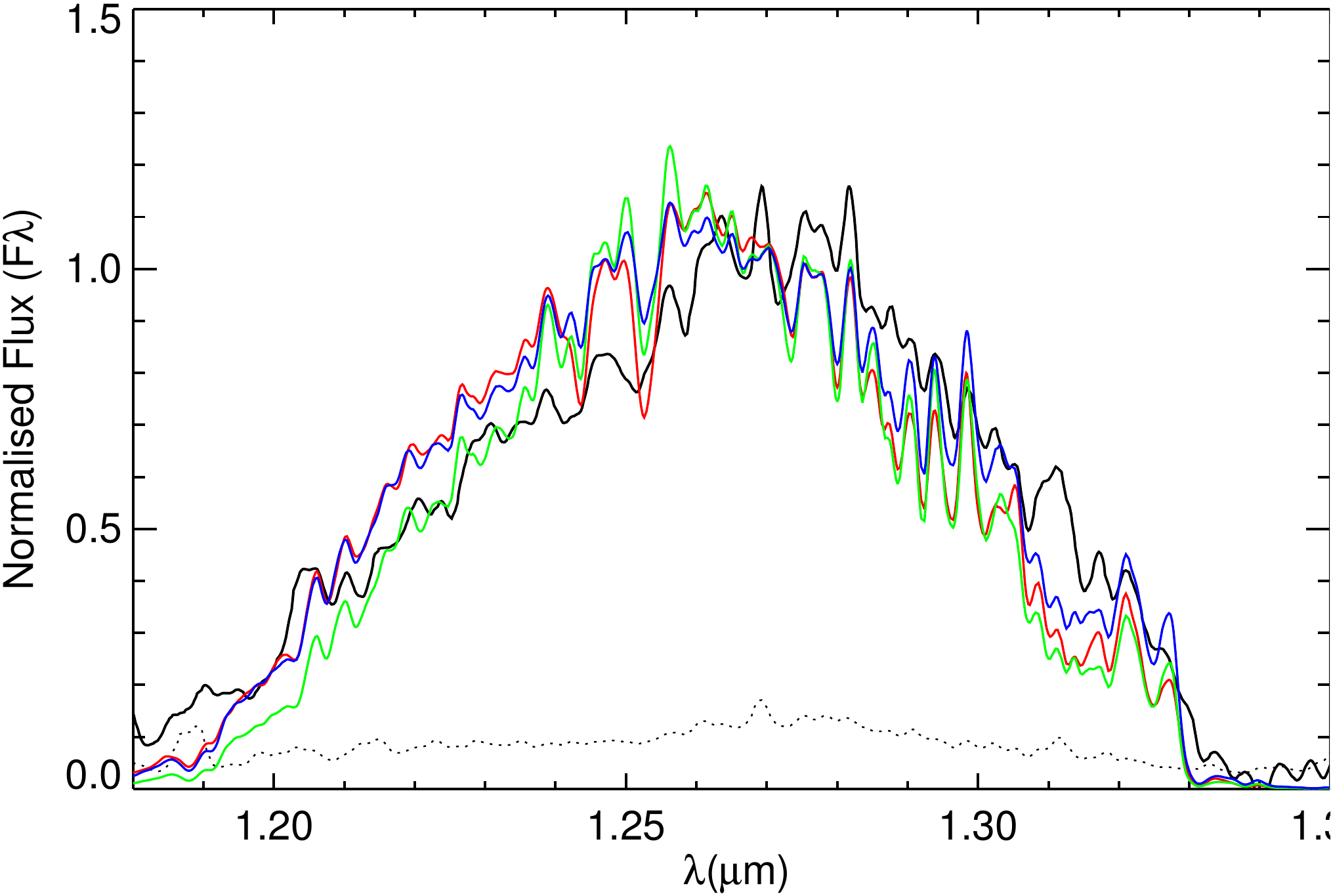}   & \includegraphics[width=9cm]{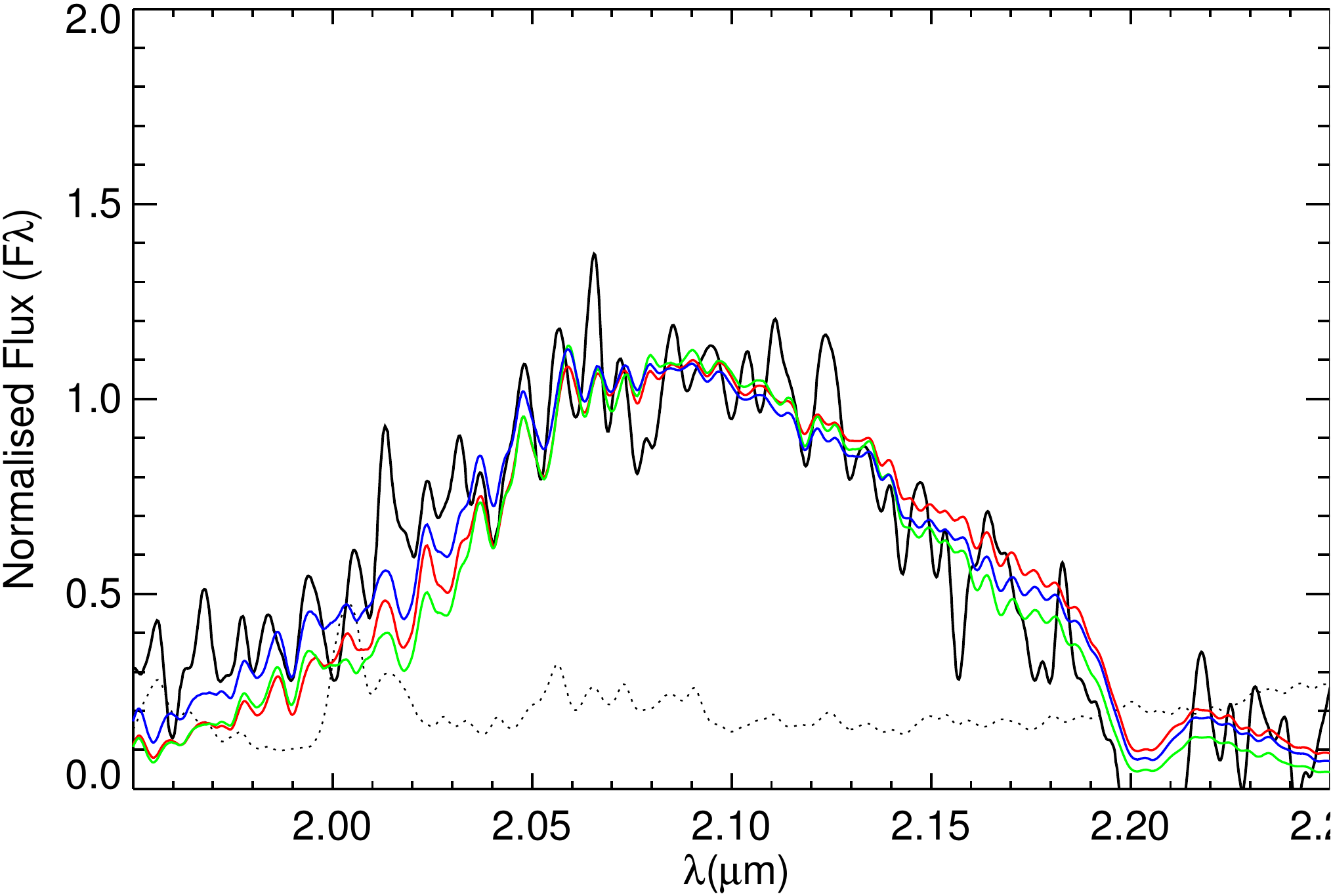}\\
\end{tabular}
 \caption{\label{bandbyband} Observed spectrum of CFBDSIR~2149 {\bf in black} in each spectral band and around the K\,{\sc i} doublet. Noise from the observed spectrum is represented by the black dotted line. From top down and left to right the panels show the spectra in $z$ at R$\sim$300, and at R$\sim$600 at all other wavelength. The spectra presented here are normalised on their local peak intensity. Best fitting BT-Settl models models in absolute flux are represented in colour.
{\bf Blue:} Model closest to the best fitting solution in absolute flux  after interpolation on both $chi ^2$ maps (700\,K, log $g$=3.5, [M/H]=0).
{\bf Red:} Model closest to the best fitting solution in absolute flux after interpolation on the high-metallicity $chi ^2$ map (800\,K, log $g$=5.0, [M/H]=0.3). 
{\bf Green:} Best fitting model in absolute flux without interpolation (650\,K, log $g$=3.5, [M/H]=0.3).}
\end{figure*}

  The overall {\bf ($\chi^2$} minima of the super-solar metallicity grid  after interpolation ($\chi^2$=1.98) is not itself at low gravity, but is located at 780~K for a gravity of log $g$=5.0. However, an important point to notice is that the super-solar metallicity $\chi^2$ surface has several local minima, corresponding to several best fit solutions whose match to the data are good. This is caused by a degeneracy in Teff/logg when matching absolute fluxes: the same absolute flux can be matched by a larger radius (smaller log $g$) and a cooler effective temperature or by a smaller radius (larger log $g$) and warmer effective temperature. Since the absolute flux is a much steeper function of effective temperature than gravity, the temperature range than can match the absolute flux is about 150~K wide, from 650 to 800~K, while the corresponding matching log $g$ range spans almost two decades. This degeneracy in the model fitting is less marked for the solar metallicity models because the shape of the observed spectrum cannot be correctly matched at higher gravity, mostly because of the significant flux excess in the observed spectra in $K$. Even though the $K$-band data has a relatively low SNR, and a correspondingly lower weight in the fit, the discrepancy between the $K$-band flux of  higher gravity models at solar metallicity and data is strong enough that higher gravity models are disfavoured and there is a global minima in the $\chi^2$ surface for low gravity. In the case of the supersolar metallicity models, metallicity enhancement increases the opacity, which causes a higher altitude, lower pressure photosphere that can match the atypical SED of \obj in $K$-band even at higher gravity. Best fitting models at high-metallicity therefore correspond to a wide range of objects, from a young isolated planetary-mass object of a few \Mjup, to moderately old brown dwars of a few tens of \Mjup. Since the local minima of the high-metallicity fit at low-gravity basically describes the same type of object as the best fitting object at solar metallicity, a young planetary mass object,  we will in the following consider mainly the best fit solution at high-metallicity, as a 800~K, log $g$=5.0 intermediate mass brown dwarf, and the best fitting solution at solar metallicity as a 700~K, log $g$=3.5, young planetary mass object, as shown on Fig. \ref{fullrange_absolute}. This allows us to investigate a truly distinct alternative when discussing the nature of CFBDSIR~2149. \\

 Although such a comparison with BT-Settl atmosphere models suggests that \obj is not a field gravity object at solar metallicity, it cannot discriminate by itself between a low gravity, solar metallicity atmosphere and a metal-enriched atmosphere. Finally we note that the main discrepancies in absolute flux between the low-gravity model and the observations are a significant lack of observed flux in the $J$ band and even more so in the $Y$ band, which could be similar to the reddening observed for unusually red L~dwarfs. This spectral peculiarity has been proposed to be linked to the presence of a high altitude dust haze in their atmosphere by \citet{Hiranaka.2012,Hiranaka.2016} and \citet{Marocco.2014}.  We note that interstellar extinction in the direction of our target \citep[Av=0.095~mag, integrated on the full line of sight][]{Schlegel.1998} cannot cause a noticeable reddening. However, such a reddening could also be caused by the presence of an inversion layer in the upper atmosphere, as proposed by \citet{Tremblin.2015}, who manage a good fit to the spectra of ROSS458C, which is quite similar to the spectra of \obj according to \citet{Delorme.2012a}. However, we note that Fig. 1 of \citet{Tremblin.2015} shows that super solar metallicity is necessary to match the very red $J-K$ colour of ROSS458C. It appears the out of equilibrium chemical processes  put forward by \citet{Tremblin.2015} have only a minor impact on the $K$-band flux of late T objects and therefore cannot explain alone the atypical $K$-band flux of ROSS458C and \obj.

 \subsection{Potassium lines equivalent width}
 Since the effects of low gravity and high metallicity on the global SED of cool atmospheres are very similar, finding an observable spectral feature that respond differently to a variation of gravity or metallicity would be key to break the degeneracy between these two physical parameters. Potassium lines are in theory a very powerful proxy to investigate metallicity in cool atmospheres, because metallicity enrichment means more potassium in the atmosphere and results into strongly enhanced K\,{\sc i} lines at constant effective temperature.
   We derived the equivalent widths of the K\,{\sc i} lines around 1.25~$\mu$m using the prescriptions detailed in \citet{McLean.2003} and \citet{Faherty.2014}.
   We did not study the K\,{\sc i} doublet around 1.17~$\mu$m because there is almost no signal left in the spectra of late-T dwarfs in this area of high water absorption. For \obj we retrieved equivalent widths of  1.4$\pm$0.6$\AA$ and 3.5$\pm$0.6$\AA$ for each component of the $J$-band doublet, comparable to the equivalent widths found by \citet{McLean.2003} using the same formalism for Gl~570D (T7.5) and 2M0415 (T8), see Table \ref{linewidth}. We also derived  equivalent widths for the BT-Settl 2014 models that produced the best fit of the overall spectrum of CFBDSIR~2149 (see section \ref{secspectro}). Though the K\,{\sc i} lines contribute negligibly to the overall $\chi^2$ in the fit, the best fitting model with very low gravity (log $g$=3.5) leads to the best agreement by far in K\,{\sc i} equivalent widths.\\

 The high gravity model that provides a relatively poor fit to the overall SED also has K\,{\sc i} equivalent widths that are much higher than we observe. This is because the correspondingly higher effective temperature of the model leads to stronger K\,{\sc i} absorption bands, as corroborated by observations. It is to be noted that the field-gravity late-T dwarfs shown on table \ref{linewidth} have lower temperatures than the best fitting high gravity models here. While these  objects (as well as cooler high gravity models) have K\,{\sc i} equivalent widths in reasonable agreement with those of CFBDSIR~2149, it has already been shown in \citet{Delorme.2012a} that their overall spectra is strikingly different from CFBDSIR~2149. They both have a lower flux in $H$-band and even more in $K$-band, a difference that would arise naturally from increase CIA absorption if these field objects have higher surface gravity than \obj. In this respect it appears that field-gravity solutions can either roughly match the spectra in $Y,J$ and $H$ bands but significantly miss the observed potassium line width or match the lines and significantly miss the overall spectra.\\
 The high metallicity model at 800~K, which fits  the overall SED of CFBDSIR~2149  including its strong $K$-band flux, completely fails at reproducing the observed K\,{\sc i} equivalent widths, both because of its higher effective temperature and because of its higher metallicity. In fact, even for the low effective temperature (650\,K), high metallicity best fit, the equivalent widths of K\,{\sc i} lines are still significantly higher than the observed values. These findings are not in contradiction -nor do they support- the analysis by \cite{Knapp.2004}, that claims that low gravity increases the K\,{\sc i} equivalent width. Indeed the low gravity best fit to \obj spectrum has a smaller K\,{\sc i} equivalent width, because it also has a lower effective temperature than the high gravity best fit, not because of its lower gravity. It appears the effect on K\,{\sc i} lines of a 100~K change in effective temperature or of 0.3~dex in metallicity  is much stronger than any effect a decade or more of change in gravity at constant effective temperature could cause.
If this modeled trend is confirmed by future observations of benchmark objects of known metallicity, this would provide an efficient way to discriminate low gravity atmospheres from high metallicity atmospheres, which otherwise present very similar spectral features in the NIR. Such a diagnostic would be critical in the study of imaged massive exoplanets, which are low gravity and could also be metal-enriched depending on their formation. In the case of CFBDSIR~2149, the  observed weak equivalent widths of the 1.25~$\mu$m  K\,{\sc i} doublet  favours the hypothesis of a solar-metallicity, low-gravity object.

\begin{table}
\caption{ Equivalent width (in $\AA$) of potassium lines at 1.243 and 1.254 $\mu$m\label{linewidth}}
\begin{tabular}{|l|c c|} \hline 
      Object name or model parameters  & Line 1  &  line 2 \\
          (log $g$/T$_{Eff}$/[M/H])  &   &  \\ \hline  \hline
  CFBDSIR~2149 (T7.5p) & 1.4$\pm$0.6&3.5$\pm$0.6\\ \hline 
    Gl~570~D$^1$ (T7.5) & 1.7$\pm$0.6&2.6$\pm$0.8\\ \hline
   2M0415$^1$ (T8) & 1.0$\pm$0.7&1.8$\pm$0.9\\ \hline 
    3.5/700~K/[0.0]  $^2$   &1.36 &3.52\\ \hline 
    5.0/850~K/[0.0]  $^3$    &4.26 &7.94\\ \hline 
    4.5/800~K/[+0.3] $^4$  &6.33 &10.84\\ \hline 
    3.5/650~K/[+0.3] $^5$  & 2.36 & 5.60\\ \hline 
    5.0/800~K/[+0.3] $^6$   &5.84 &10.56\\ \hline 
\end{tabular}
\tablefoot{$^1$From \citet{McLean.2003}. $^2$Best fitting  BT-Settl model at low gravity
  $^3$Best fitting model at high gravity after flux normalisation 
 $^4$Best fitting model at high metallicity after flux normalisation 
 $^5$Best fitting model in absolute flux 
 $^6$Best fitting model in absolute flux at high metallicity and high gravity}
\end{table}

\subsection{Spectral Synthesis: exploring different atmosphere models}

{To explore whether our spectral synthesis conclusions were strongly dependent on the set of models used, we fitted our observational data to other available state-of-the-art models, starting with those of \citet{Burrows.2003}. This model grid has 32 spectra with effective temperatures and gravities self-consistently derived from evolutionary models \citep{Burrows.1997} for objects ranging from 1 to 25~\Mjup and ages ranging from 100~Myr to 5~Gyr. The \citet{Burrows.2003} model grid does not include super-solar metallicity objects, and it is relatively coarse so that it is possible that the real $\chi ^2$ minimum could be relatively far from any node of the model grid. However, it is interesting to note that the best fit ($\chi ^2$=3.4) corresponds to a 7~\Mjup free-floating planet aged 100~Myr, very similar to our BT-Settl results.\\

In order to make a more detailed comparison, we compare the data to atmosphere models similar to those described in \citet{Morley.2012,Morley.2014}, which include opacities for sulfide and salt clouds that condense in T and Y dwarfs. The models we use here include two major updates to the opacities, including a new methane line list \citep{Yurchenko.2014} and alkali line list \citep{Allard.2005}. Chemical equilibrium calculations \citep{Lodders.2006book,Visscher.2010} have also been revised and extended to include higher metallicities. These updates will be described in detail in a set of upcoming papers that focus on the new model grid (Marley et al. in prep., Morley et al. in prep., hereafter Ma\&Mo2017). 

 Unlike BT-Settl, the Ma\&Mo2017 model grid is not coupled with evolutionary models, so we cannot directly fit to our spectrum in absolute flux because there is no radius associated with a given atmosphere. The fits are therefore carried out after a normalisation in flux as described in section 4.1.1. The resulting normalisation factor physically correspond to assigning a radius to the object, by scaling the flux of the model to the observed absolute flux of the object. Then, knowing the surface gravity of the model and the radius of the object, the mass of the object can be determined by applying Newton's law. However, we adapted our fitting procedure so that it derives the radius corresponding to the flux (and gravity and mass) of the model. The only other modification of our fitting procedure with respect to what we used for the BT-Settl models is that we kept the $H$-band peak in the fit because the Ma\&Mo2017 models use the latest methane line list, which enables a good fit in the $H$ band. The grid covers temperatures from 450 to 900~K in steps of 25~K, with gravity ranging from 3.5 to 5.5~dex in steps of 0.5~dex and metallicities ranging from 0.0 to 1.0 dex in steps of 0.5 dex. Thus, the Ma\&Mo2017 grid explores higher metallicities than the BT-Settl grid.

The overall best fitting model is a 900~K, log $g$ =5.5, solar metallicity atmosphere ($\chi ^2$=1.71). Although this atmosphere completely fails at reproducing the $Y$-band peak and the $J$-band potassium doublet, it fits most of the data very well, notably in the high signal-to-noise part of the spectrum that has  more weight in the $\chi ^2$ calculation. However, this atmosphere would correspond to a 51~\Mjup brown dwarf with a radius of only 0.66~\Rjup. According to the Cond evolutionary models of \citet{Baraffe.2003}, a 50~\Mjup brown dwarf is expected to reach a temperature of about 900~K after 5~Gyr, with an expected radius of 0.83~\Rjup that is much larger than the best-fit radius of 0.66~\Rjup.  In fact, the radius that fits the observed flux of \obj for this high-temperature model is also well below the minimum possible radius for any brown dwarf, even for an age of 10~Gyr the radius only reaches 0.77~\Rjup for a 72~\Mjup object. The minimum theoretical radius is governed by the well established physics of electron degeneracy pressure \citep{Kumar.1963}, so we reject this high-temperature, small-radius solution as unphysical. We note that the best-fit atmosphere with 900~K and log $g$=5.5 would have a physically plausible radius if the flux were hypothetically higher due to a larger distance of at least 68~pc.  This distance would correspond to a parallax of 14.7~mas, more than 2$\sigma$ smaller than our measured parallax of 18.3$\pm$1.8~mas.

If we remove the highest gravity log $g$=5.5~dex models from consideration, the best fitting overall solution is a 775~K, log $g$=4.5~dex solar metallicity atmosphere ($\chi ^2$=1.96).  This model provides a much better fit in the $Y$ and $H$ bands (Fig. \ref{model_morley}), but it is not as good in the $J$ band as the previous solution. The radius and mass implied by this atmosphere are 13~\Mjup and 1.1~\Rjup, which almost exactly matches the substellar evolution models at 500~Myr for a 13~\Mjup object.

If we force super-solar metallicity, we find the best fit is a 800~K, log $g$=5.0~dex atmosphere ($\chi ^2$=2.22) with moderately high metallicity, [M/H]=+0.5~dex, very similar to the best-fit solution of the BT-Settl model grid. The radius and mass corresponding to this atmosphere are 0.87~\Rjup and 35~\Mjup, in good agreement with Cond evolutionary model predictions for a 30-40~\Mjup object at an age of 3~Gyr that would have a radius of about 0.9~\Rjup. It is interesting to note that although the Ma\&Mo2017 model grid does allow for a larger maximum metallicity of [M/H]=+1.0~dex, the best  {\bf high metallicity} fit to our data is achieved using the intermediate step of [M/H]=+0.5~dex.

Finally, forcing low gravity solutions also provides a good fit to the data (log $g$=4.0, T=775~K, $\chi ^2$=2.26), associated with the a radius of 1.06~\Rjup and a mass of 4.4~\Mjup. There is some tension with evolutionary model predictions that forecast that a 4-5~\Mjup objects cools down to 750-800~K  for an age  40-50~Myr, with a larger radius of $\sim$1.25~\Rjup.  This would require \obj to be somewhat farther away, at a distance of  $\sim$65~pc and 1.6$\sigma$ different from our parallax measurement, in order to be self-consistent. Though this low gravity solution is only marginally compatible with the measured absolute flux of the object, the continuity with the best overall fit, with the same effective temperature and log $g$=4.5, hints that a denser sampling in the models lo g grid could have revealed physically self-consistent solutions with good fir to the data for gravity between log $g$=4.0 and log $g$=4.5. This range in gravity corresponds to masses between 5 and 13~\Mjup. We also note that all of our fits converged toward either the fastest dust sedimentation speed (thinnest clouds) or to the \textit{no cloud} setting (explicitly removing all dust from the atmosphere). Therefore, Ma\&Mo2017 models strongly favour clear over dusty atmospheres. \\

 After exploring model grids independent from BT-Settl, we conclude that the best-fit  Ma\&Mo2017 models are quite similar to those from BT-Settl. Two families of self-consistent solutions emerge, with the high metallicity, high gravity one being almost identical between the two model grids, having a temperature of 800~K, a gravity of log $g$=5.0~dex and a moderate metallicity. The second family of plausible solutions indicates a relatively young planetary mass object. {\bf However} the BT-Settl grid points toward cooler and lower gravity solutions, corresponding to objects well within the planetary mass range, while the Ma\&Mo2017 models are consistent with objects slightly warmer, older and more massive, around 13~\Mjup. Because of this higher effective temperatures none of these latter models can reproduce the very weak observed K\,{\sc i} doublet at 1.25~$\mu$m (Fig. \ref{model_morley_details}). Thus, the potassium doublet would tend to favor the lower temperature found by the BT-Settl models, but on the other hand we note that the updated methane line list used by the Ma\&Mo2017 models strongly improves the quality of the fit in the $J$ and $H$-band peaks, possibly making this latter model grid more reliable. It is encouraging to note that all the best-fitting Ma\&Mo2017 models detailed in this section also correspond to local minima in the BT-Settl $\chi ^2$ maps, so perhaps the small differences between the two sets of best-fitting models arise partly from the different grid sampling rather than from different physics.

%

\begin{figure*}
\includegraphics[width=18cm]{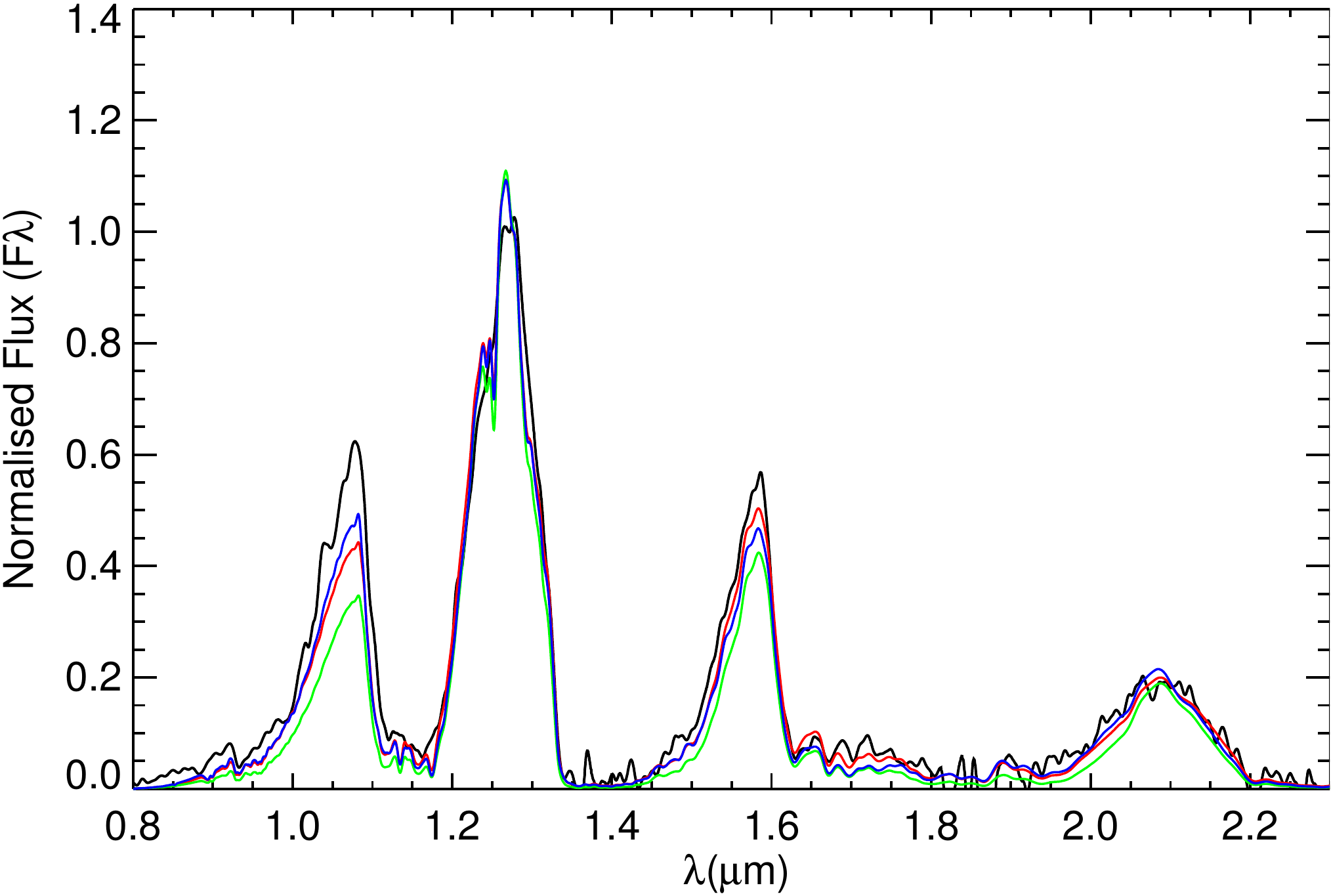}
\caption{\label{model_morley} Observed spectrum of CFBDSIR~2149 {\bf in black} in each spectral band. Best fitting Ma\&Mo2017 models after normalisation on the $J$ band peak are represented in colour. The spectra presented here have been binned to R$\sim$200.
{\bf Blue:} Best fitting Ma\&Mo2017 low-gravity model  (775\,K, log $g$=4.0, [M/H]=0)
{\bf Red:} Best fitting Ma\&Mo2017 model (775\,K, log $g$=4.5, [M/H]=0).
{\bf Green:} Best fitting Ma\&Mo2017 high-metallicity model  (800\,K, log $g$=5.0, [M/H]=+0.5). }
\end{figure*}

\begin{figure*}
\begin{tabular}{cc} \\
\includegraphics[width=9cm]{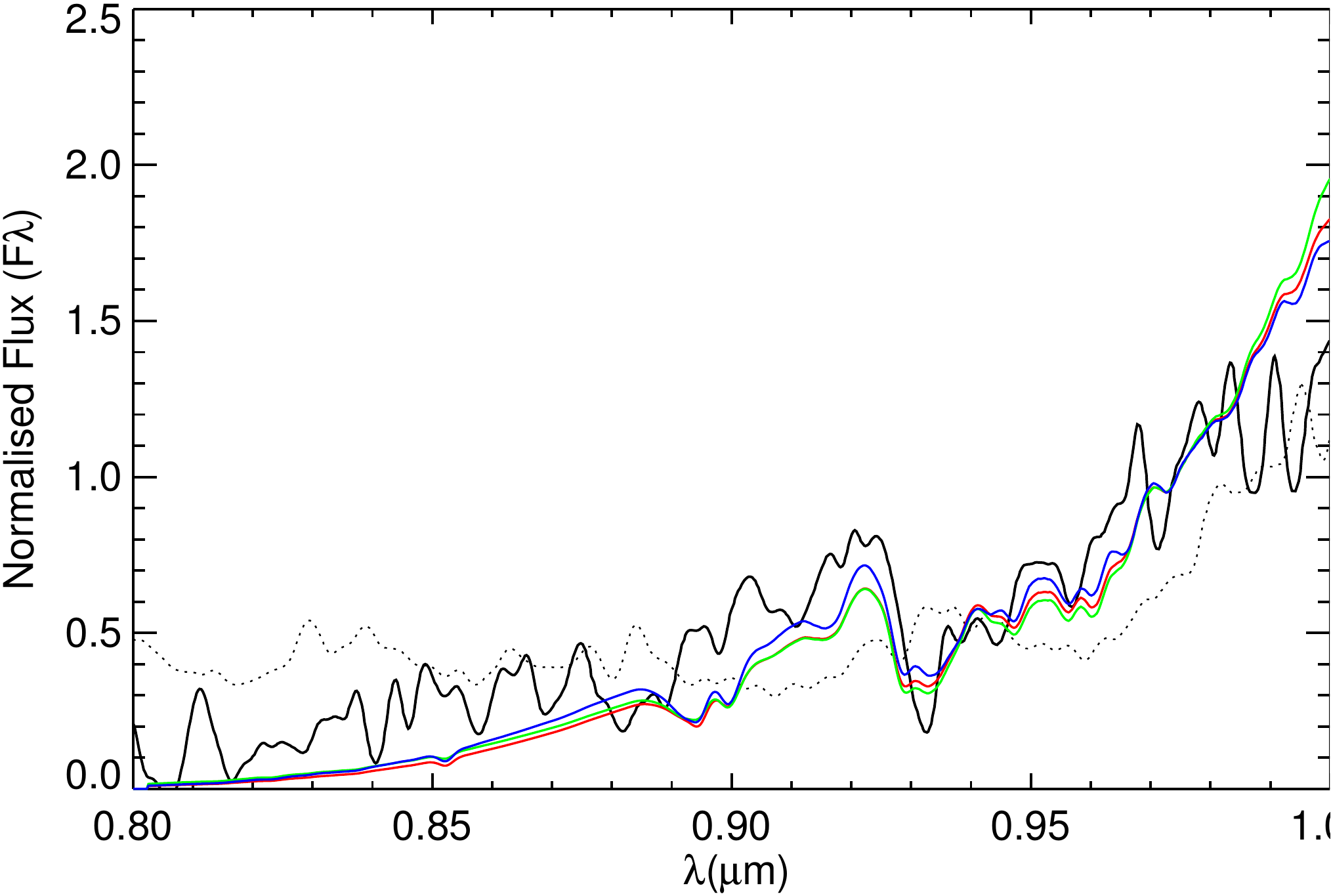} & \includegraphics[width=9cm]{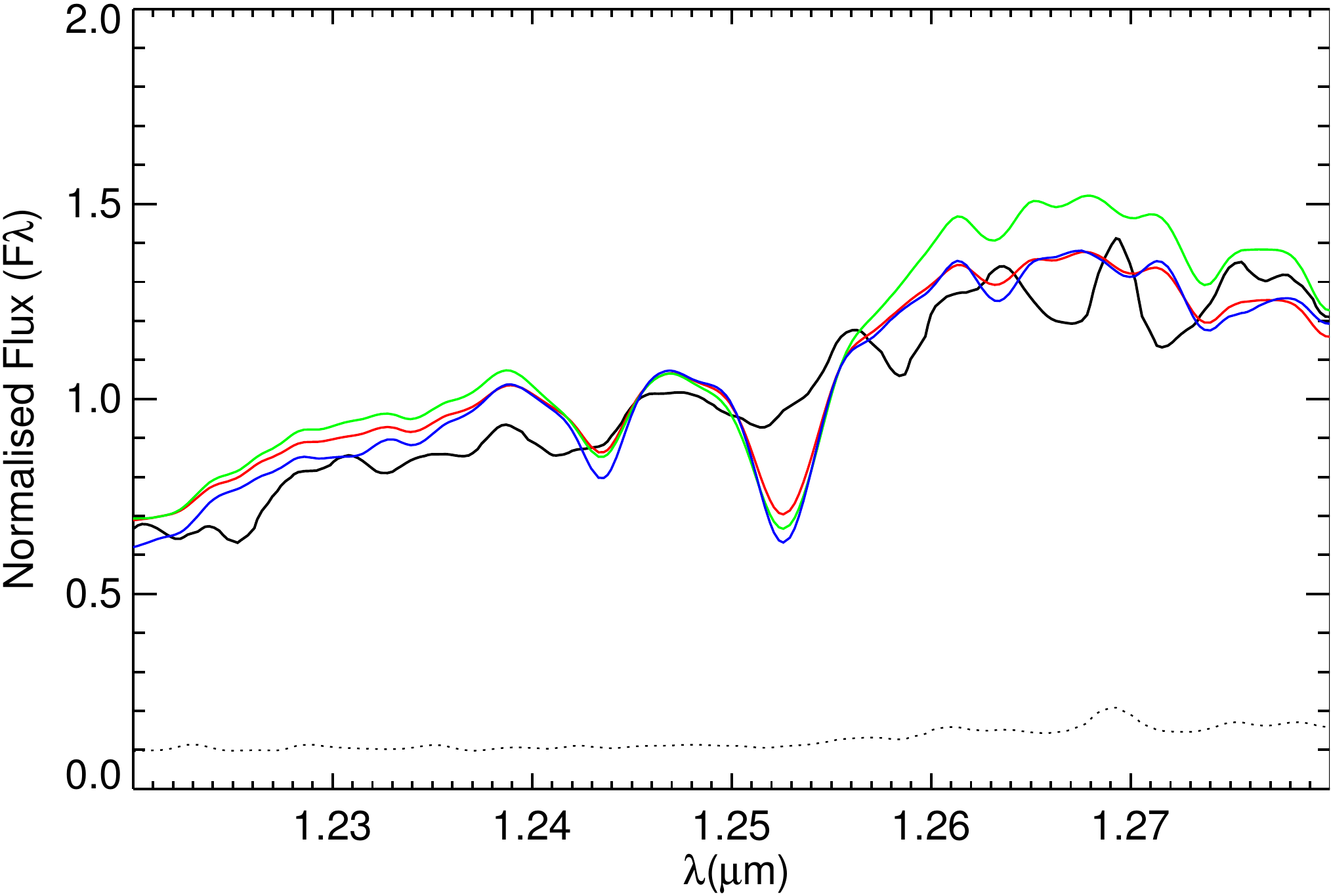} \\
\includegraphics[width=9cm]{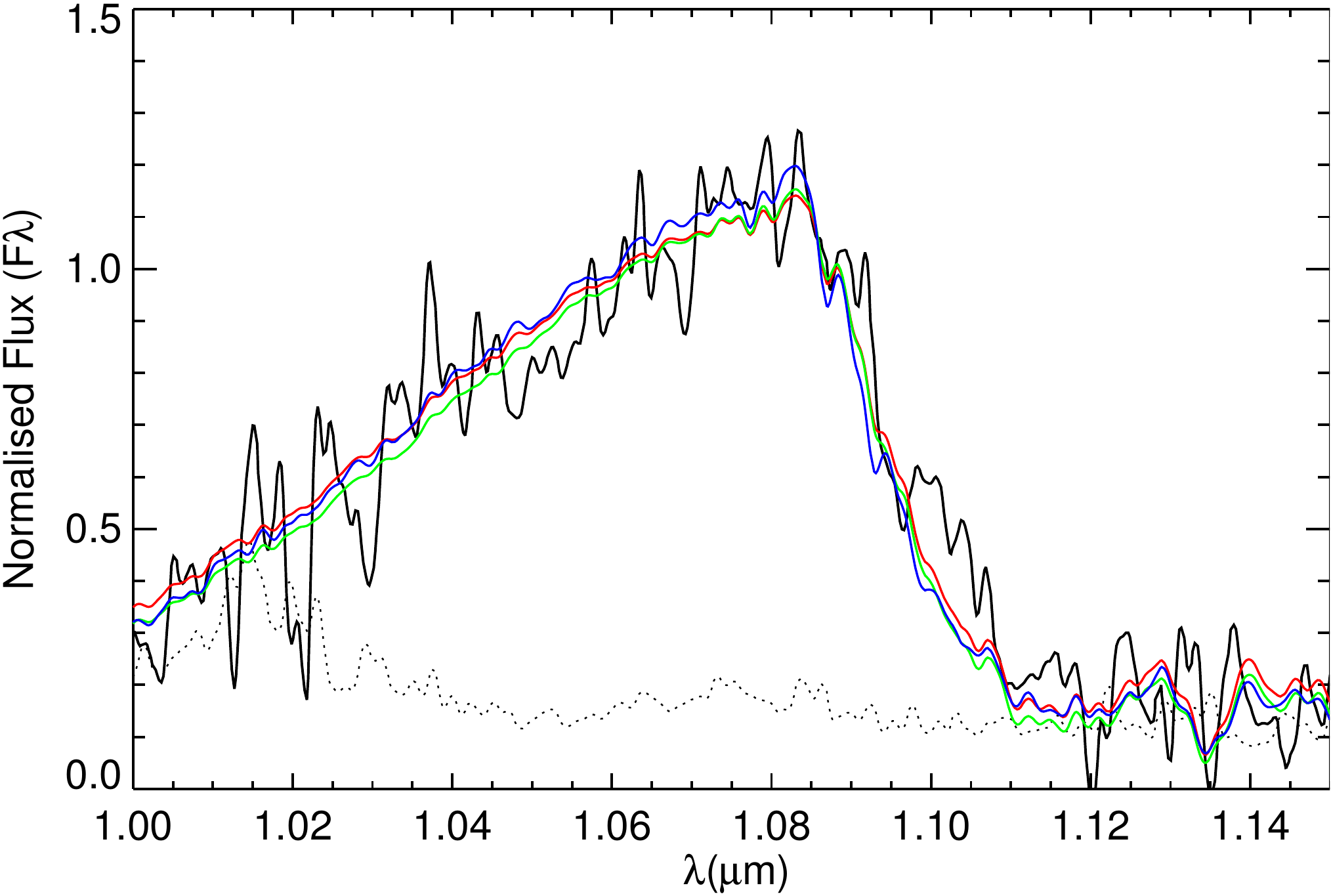}   & \includegraphics[width=9cm]{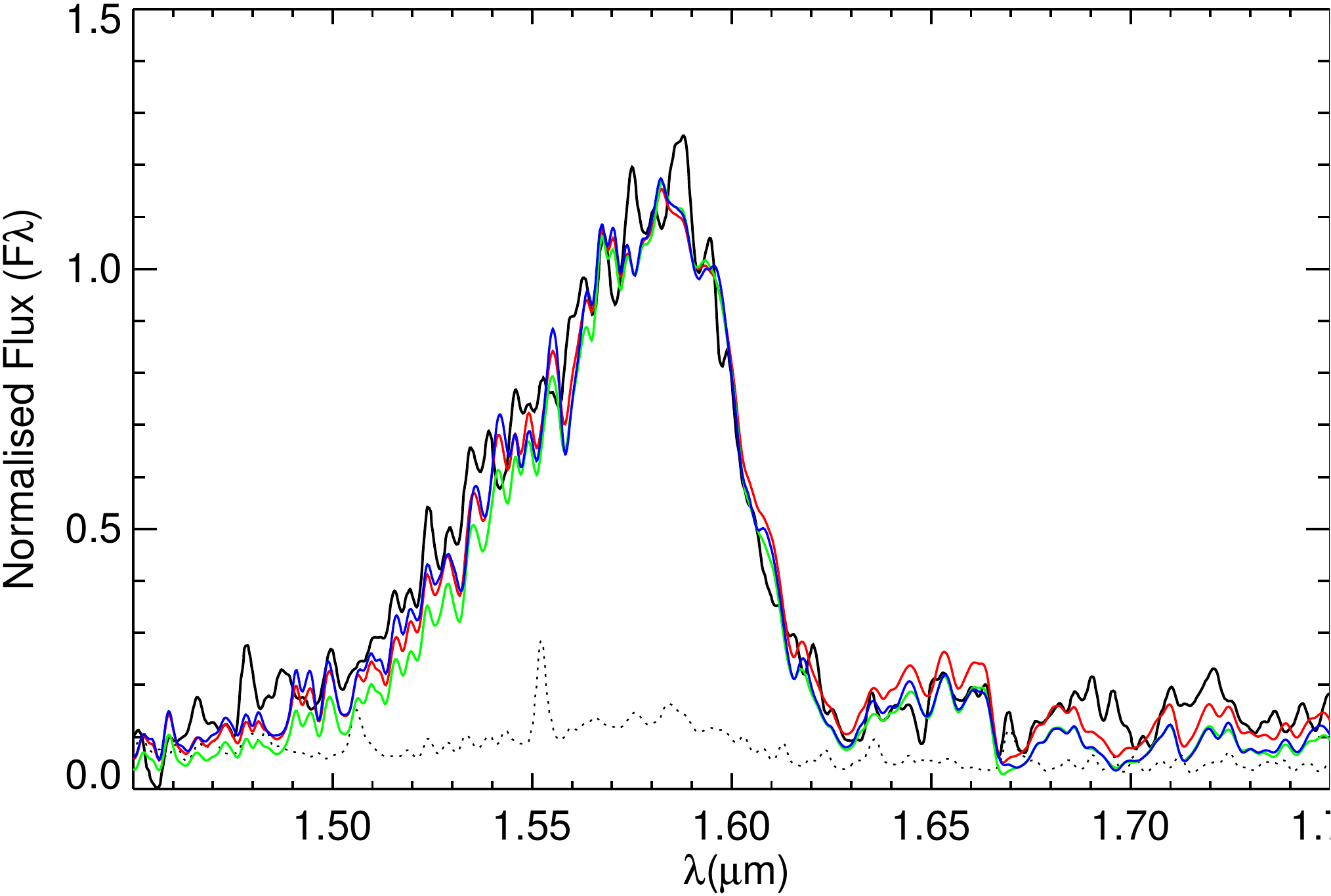}\\
\includegraphics[width=9cm]{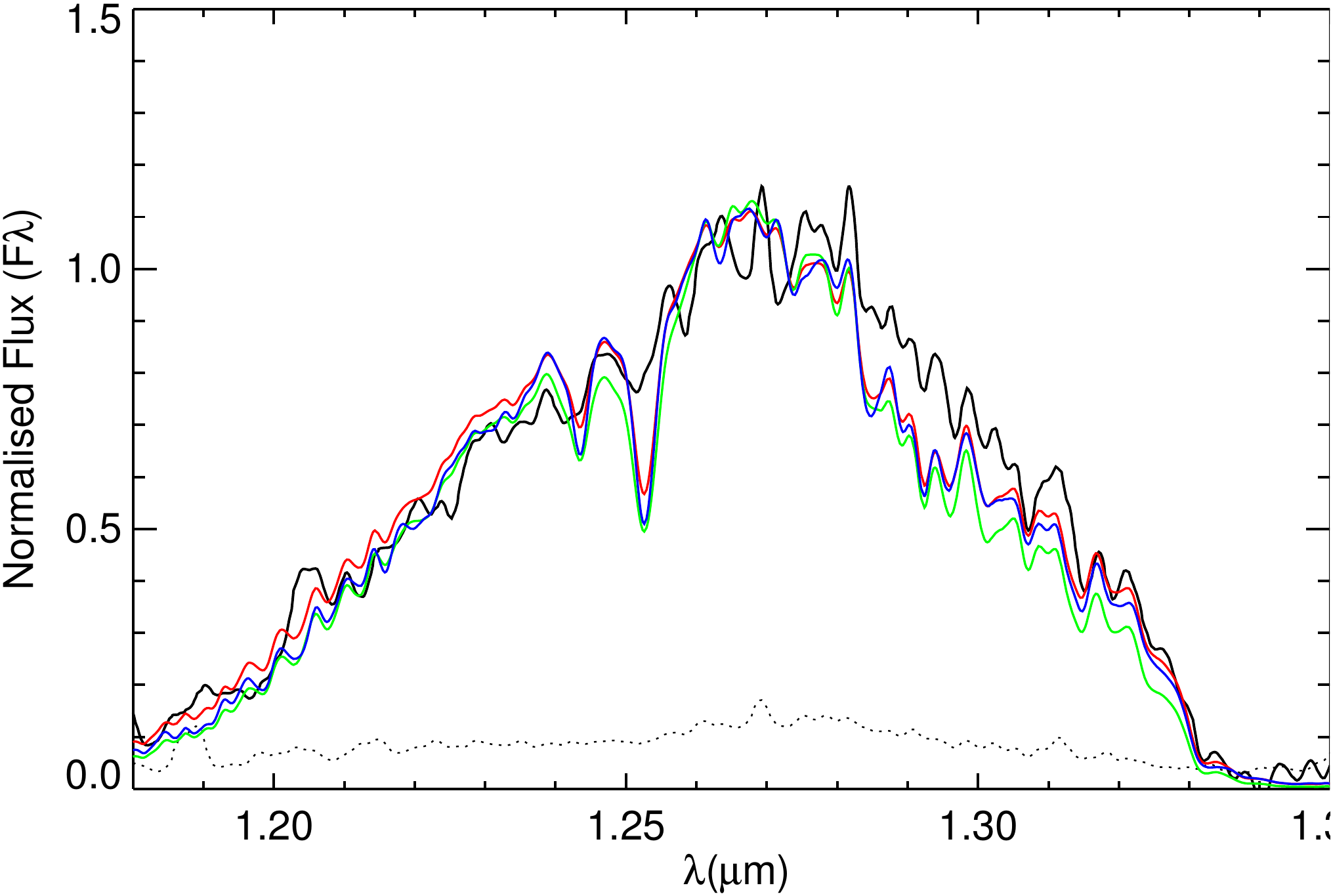}   & \includegraphics[width=9cm]{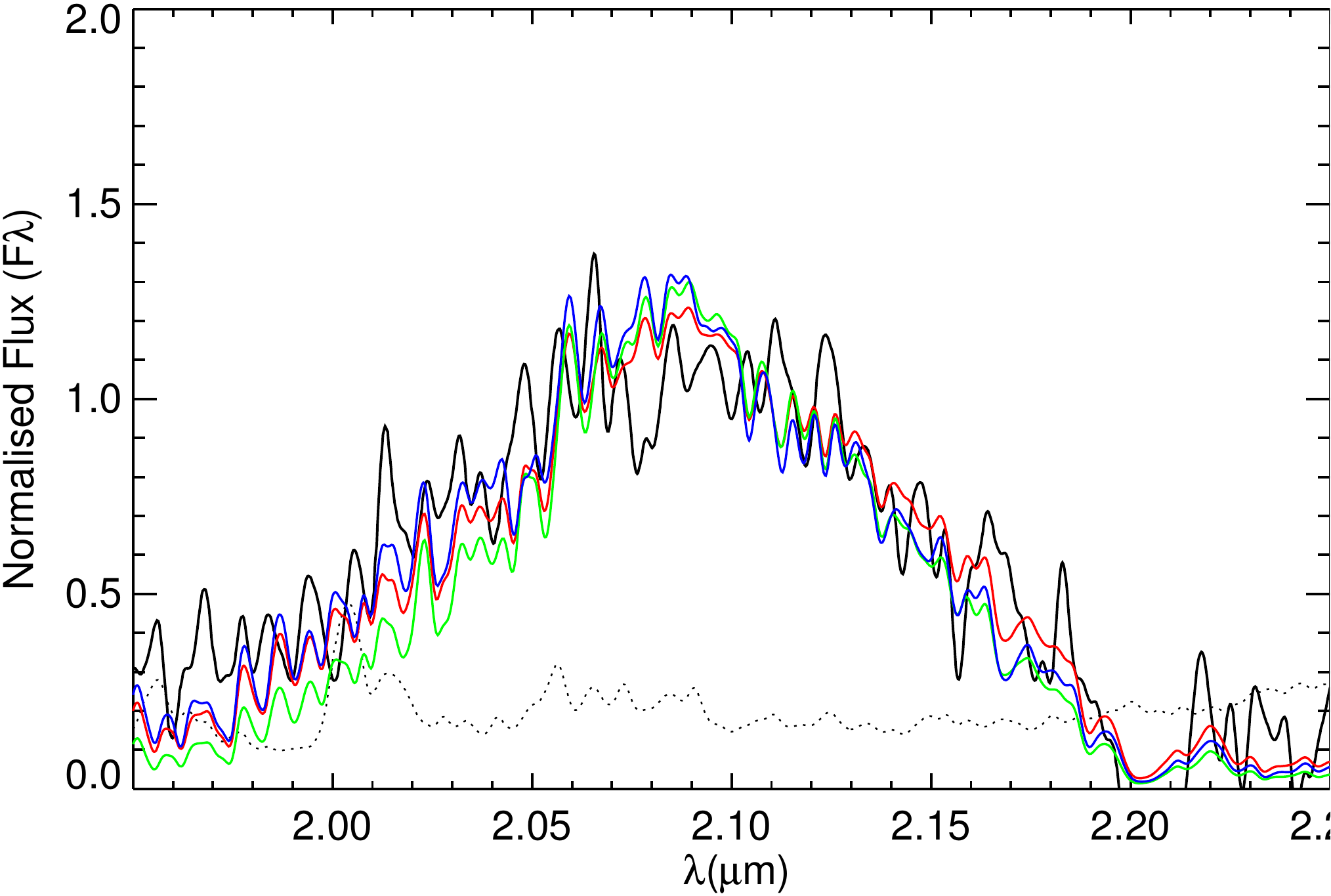}\\
\end{tabular}
 \caption{\label{model_morley_details} Observed spectrum of CFBDSIR~2149 {\bf in black} in each spectral band and around the K\,{\sc i} doublet. Best fitting Ma\&Mo2017 models after normalisation on the $J$ band peak are represented in colour. Noise from the observed spectrum is represented by the black dotted line. From top down and left to right the panels show the spectra in $z$ at R$\sim$300, and at R$\sim$600 at all other wavelength. The spectra presented here are normalised on their local peak intensity.
{\bf Blue:} Best fitting Ma\&Mo2017 low-gravity model  (775\,K, log $g$=4.0, [M/H]=0)
{\bf Red:} Best fitting Ma\&Mo2017 model (775\,K, log $g$=4.5, [M/H]=0).
{\bf Green:} Best fitting Ma\&Mo2017 high-metallicity model  (800\,K, log $g$=5.0, [M/H]=+0.5). }
\end{figure*}

\subsection{Comparison with a known high metallicity brown dwarf: GJ~758B}
GJ~758B \citep{Thalmann.2009}, GJ~504b \citep{Kuzuhara.2013} and Ross~458C \citep{Burningham.2011b} are the only known late T-dwarfs with probable supersolar metallicity, making them prime benchmarks to compare to CFBDSIR~2149. For all of these objects, other known late-T dwarf spectra and model spectra at high gravity and solar metallicity provide very poor fits to their atypical photometry. Since the age of GJ~504b is still unclear and its effective temperature is significantly lower \citep[500--550\,K][]{Skemer.2016} than CFBDSIR~2149, and since Ross~458C has already been compared to \obj in \citet{Delorme.2012a}, we focus our comparison on GJ~758B.  The primary GJ~758 is older than 600~Myr and has a metallicity [Fe/H]=0.18$\pm$0.05 \citep{Vigan.2016}. In Figure \ref{758b} and Table \ref{758_phot}, we show normalised narrow-band  photometry of GJ~758B \citep[T8,][]{Vigan.2016} together with the  corresponding spectrophotometry of CFBDSIR~2149. We see that the colours of GJ~758B agrees well with those of CFBDSIR~2149 in the $J$ and $H$ bands, as well as with the Spitzer photometry at 4~$\mu$m, confirming the very similar spectral types of these objects. The photometry of GJ~758B differs from \obj's spectra in the $K$ band, hinting that moderately high metallicity alone cannot reproduce the very red SED of CFBDSIR~2149. Another discrepancy is visible in the $Y$ band, with CFBDSIR~2149 being significantly bluer in $Y-J$, perhaps indicating that metallicity enhancement (by strenghtening the  alkali absorption) and high gravity (by extending the pressure-broadened wings of the very strong 0.77~$\mu$m potassium doublet into the $Y$ band) tend to blunt the $Y$-band flux in GJ~758B. It is striking that the SED of GJ~758B is distinct from that of \obj in ways that are qualitatively compatible with \obj having lower gravity than GJ~758B. Though these objects have colours that agree within 2$\sigma$ they have significantly different absolute luminosities, see table~\ref{758_phot}, with \obj being approximately five times brighter than GJ~758B. This large difference could only be explained by a combination of factors, such as a larger radius for CFBDSIR~2149, which would be consistent with low-gravity and a slightly higher effective temperature. It also opens the possibility that \obj could be an unresolved equal mass binary, but it also has to be noted that the flux of GJ~758B is highly unusual by itself \citep{Vigan.2016}.
Such a comparison between only 2 objects cannot be conclusive by itself, but it still tends to show that unusually low-gravity might be at least partly responsible for the atypical spectral features of CFBDSIR~2149. 

\begin{figure}
 \caption{\label{758b} Low resolution spectra (R=400) of \obj (black) compared with the narrow band photometry of GJ~758B from\citet{Vigan.2016}, red crosses.  Both data sets are normalised on their average flux in the $J3$ filter.}
\includegraphics[width=9cm]{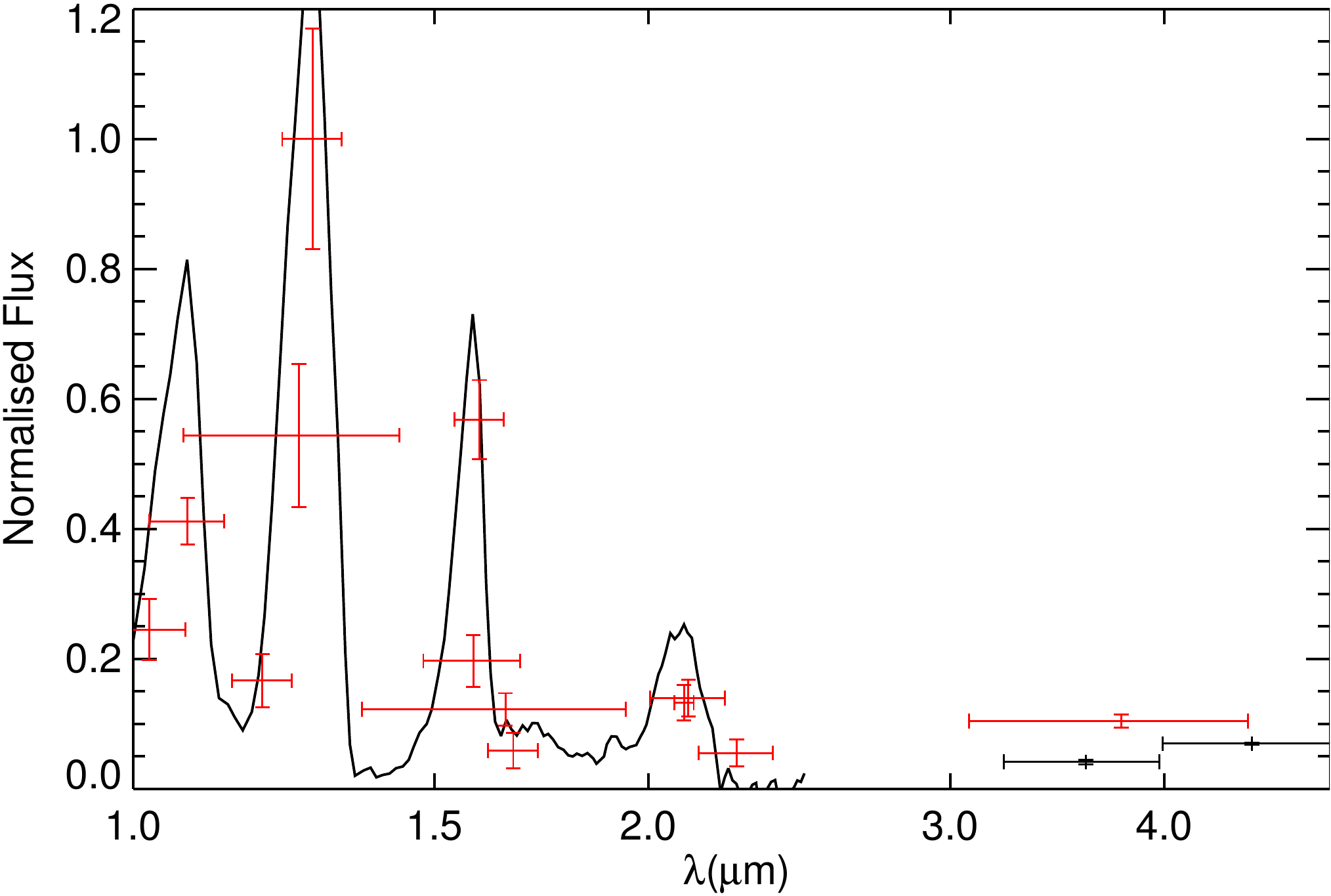}

\end{figure}

\begin{table*}
\caption{Absolute magnitudes of CFBDSIR~2149 (synthesised from spectra) and Gl758B (observed ), in the SPHERE narrow band filters. \label{758_phot}}
\begin{tabular}{|l|c c c c c c c c|} \hline 
      &  $M(Y2)$           & $M(Y3)$           &  $M(J2)$          &  $M(J3)$          &  $M(H2) $          &  $M(H3) $          &  $M(K1)$          &  $M(K2)$          \\ \hline \hline 
\obj  &  17.22$\pm$0.05 & 16.64$\pm$0.04 & 17.09$\pm$0.05 & 15.42$\pm$0.04 & 15.46$\pm$0.04  & 17.03$\pm$0.05  & 15.35$\pm$0.04 & 16.80$\pm$0.05 \\ \hline 
GJ758b & 19.19$\pm$0.20 & 18.43$\pm$0.10 & 19.06$\pm$0.25 & 16.83$\pm$0.18 & 16.59$\pm$0.12  & 18.88$\pm$0.42  & 17.03$\pm$0.21 & 17.78$\pm$0.35 \\ \hline 
\end{tabular}
\tablefoot{The uncertainty on the parallax of \obj corresponds to an additional $\pm$0.2~mag systematic error. }
\end{table*}

\subsection{Bolometric flux \label{secabsmag}}
We converted the observed absolute fluxes into bolometric flux using
the ``super-magnitude'' method described in \citet{Dupuy.2013}.  In
this approach, we combine fluxes calculated from $J$-, $H$-, $[3.6]$-,
and $[4.5]$-band photometry into a magnitude defined as $m_{JH12}$ by
\citet{Dupuy.2013}, and for CFBDSIR~2149 we compute $m_{JH12} =
19.292\pm0.025$\,mag.  Using the single bolometric correction of
BC$_{JH12} = 2.93\pm0.08$\,mag derived by \citet{Dupuy.2013} from the
models of \citet{Morley.2012}, we find $m_{\rm bol} = 22.22\pm0.08$\,mag.  We note that by using only these bandpasses, and
not including $K$ band, the \citet{Dupuy.2013} super-magnitude is more
robust against assumptions about clouds and surface gravity. This can be seen in Figure~S4
of \citet{Dupuy.2013} where, e.g., the scatter in model-derived
bolometric corrections at 700\,K is small and primarily driven by
surface gravity.  In addition, the $J-H$ colour of CFBDSIR~2149 is
normal compared to field brown dwarfs, while its $J-K$ colour is
anomalously red, and Figure~9 of \citet{Morley.2012} shows
that their models reproduce empirical $J-H$ colour--magnitude diagrams.
Therefore, we expect our super-magnitude approach to provide an
accurate estimate of the bolometric magnitude of CFBDSIR~2149.

Using our parallax of $18.3\pm1.8$\,mas, we compute a bolometric
 luminosity of $\log (L_{\rm bol}/L_\odot ) = -5.51^{+0.10}_{-0.09}$\,dex for
CFBDSIR~2149.  Given this luminosity and an assumption about the
radius, we can compute a corresponding effective temperature for
CFBDSIR~2149. Table \ref{Teffage} shows the  corresponding effective temperatures associated with various age and mass hypotheses, using the solar metallicity {\it Cond} evolutionary tracks from \citet{Baraffe.2003} to derive radius from mass and age. \\

\subsubsection{Comparison to BT-Settl atmosphere fits}
To further assess the physical nature of CFBDSIR~2149, we compare our
BT-Settl spectral fits to fundamental properties derived from
evolutionary models given our measured luminosity and a range of
possible ages (Table \ref{Teffage}).  The overall best fitting model spectrum (\Teff$ =650$\,K, log $g = 3.5$\,dex, [M/H]=0.3) has  $\log (Lbol/ L\odot ) = -5.42 $ from its radius and effective temperature. This agrees quite well with very young ages (20--50\,Myr), corresponding to masses of $\approx$2--5\,\Mjup. 
 The model spectrum at solar metallicity that is closest to the global $\chi^2$ minimum, (\Teff$ =700$\,K, log $g = 3.5$\,dex, [M/H]=0 ) has  $\log (Lbol/ L\odot ) = -5.29$, also agreeing best with very young ages and planetary masses. In this age range the effective temperatures expected from the bolometric luminosity are slightly lower, and the radius slightly smaller than the best fit on our model grid. However this slightly smaller radius and effective temperature are necessary to match the observed fainter bolometric luminosity of \obj  and are well within the expected uncertainty of atmosphere models. Also we note that after interpolating $\chi ^2$ between the points of the discrete grid model, the overall  $\chi ^2$ minimum consistently correspond to a lower effective temperature (\Teff$ =680$\,K), associated with a fainter object.

Evolutionary models predict that by an age of
100\,Myr the gravity reaches log $g = 4.06\pm0.04$\,dex at a mass of
$6.4\pm0.6$\,\Mjup, which is one model grid step higher than the best
fit value and thus perhaps marginally consistent.  At older ages the solutions are, as expected, increasingly discrepant with the very low gravity hypothesis, which thus implies an age younger than about 100~Myr.  The
best-fit model atmosphere parameters when restricted to high
metallicity (\Teff $ = 800$\,K, log $g = 5.0$\,dex,$\log (Lbol/ L\odot ) = -5.43$) are inconsistent
with evolutionary model parameters at young ages but are consistent for ages of a few  Gyr and corresponding masses of
20--40\,\Mjup.  At ages older than this gravity is significantly
higher, and at younger ages \Teff\ is significantly cooler, constraining the age range within which the high-metallicity, field gravity hypothesis is consistent to around less than 5~Gyr.
Therefore, we conclude that both of the scenarios are self consistent within restricted age ranges
from the perspective of evolutionary and atmosphere models. We also note that for an age around 5~Gyr the best fit for high gravity (800\,K, log $g$=5.0, [M/H]=0,$\log (L_{\rm bol}/L_\odot ) =-5.43 $), is also consistent with the measured bolometric luminosity of CFBDSIR~2149. The main issue with this field-gravity hypothesis is that it significantly fails at reproducing the observed colours and spectra of CFBDSIR~2149.

\begin{table}
\caption{Effective temperature, gravity, mass and radius derived from the absolute luminosity of CFBDSIR~2149 for various age hypotheses and solar metallicity  .\label{Teffage} }
\begin{tabular}{|l|c c c c|} \hline
 age (Gyr) & \Teff(K) & log $g$ &M.(\Mjup)&R.(R$_{\rm Jup}$) \\ \hline \hline
 0.02 & $675_{ -35}^{ +35}$ & $3.58_{-0.04}^{+0.04}$ & $2.6_{-0.2}^{+0.3}$  & 1.271$\pm$0.007 \\ \hline
 0.05 & $685_{ -40}^{ +35}$ & $3.85_{-0.05}^{+0.04}$ & $4.3_{-0.5}^{+0.4}$  & 1.231$\pm$0.001 \\ \hline
 0.1  & $695_{ -35}^{ +35}$ & $4.06_{-0.04}^{+0.04}$ & $6.4_{-0.6}^{+0.6}$  & 1.176$\pm$0.001 \\ \hline
 0.5  & $735_{ -45}^{ +30}$ & $4.42_{-0.02}^{+0.01}$ & $12.3_{-0.7}^{+0.3}$  &1.066$\pm$0.003 \\ \hline
 1.0  & $765_{ -45}^{ +45}$ & $4.74_{-0.04}^{+0.04}$ & $20.9_{-1.8}^{+1.7}$ &  0.979$\pm$0.009 \\ \hline
 5.0  & $845_{ -50}^{ +50}$ & $5.25_{-0.04}^{+0.04}$ & $47.0_{-3.0}^{+3.0}$ & 0.811$\pm$0.008  \\ \hline \hline
 BT Models & Teff(K) & log $g$ &M(\Mjup)&R(R$_{\rm Jup}$)  \\ \hline \hline
 Low-g.$^1$      & 700 & 3.5 & 2.7 & 1.53   \\ \hline
 Field 1$^2$      & 850 & 5.0 & 38 & 0.99  \\ \hline
 High-m.1$^3$     & 650 & 3.5 & 2.7 & 1.53   \\ \hline
 Field 2$^4$      & 800 & 5.0 & 38 &  0.99 \\ \hline
 High-m.2$^5$     & 800 & 5.0 & 38 &  0.99 \\ \hline\hline
 Ma\&Mo Models & Teff(K) & log $g$ &M(\Mjup)&R(R$_{\rm Jup}$)  \\ \hline \hline
 Low-g.$^6$      & 775 & 4.0 & 4.3 & 1.06   \\ \hline
 Field $^7$      & 900 & 5.5 & 51 & 0.66  \\ \hline
 High-m.$^8$     & 800 & 5.0 & 35 & 0.87   \\ \hline
 Best fit$^9$      & 775 & 4.5 & 13 &  1.07 \\ \hline
\end{tabular}
\tablefoot{$^1$Best fitting BT-Settl model after normalisation at low gravity (700\,K, log $g$=3.5, [M/H]=0,$\log (Lbol/ L\odot ) = -5.29$).  $^2$Best fitting BT-Settl model after flux normalisation at high gravity (850\,K, log $g$=5.0, [M/H]=0,$\log (Lbol/ L\odot ) =-5.32 $). $^3$Best fitting BT-Settl model in absolute flux (650\,K, log $g$=3.5, [M/H]=+0.3,$\log (Lbol/ L\odot ) = -5.42$).  $^4$Best fitting BT-Settl model in absolute flux at high gravity (800\,K, log $g$=5.0, [M/H]=0, $\log (Lbol/ L\odot ) =-5.43 $). $^5$Best fitting BT-Settl model in absolute flux at high metallicity (800\,K, log $g$=5.0, [M/H]=+0.3, $\log (Lbol/ L\odot ) =-5.43 $).
 $^6$Best fitting Ma\&Mo2017 model at low gravity (775\,K, log $g$=4.0, [M/H]=0)
 $^7$Best fitting Ma\&Mo2017 model at high gravity (900\,K, log $g$=5.5, [M/H]=0)
 $^8$Best fitting Ma\&Mo2017 model at high metallicity (800\,K, log $g$=5.0, [M/H]=+0.5)
 $^9$Best fitting overall Ma\&Mo2017 model (775\,K, log $g$=4.5, [M/H]=0)}
\end{table}

\subsubsection{Comparison to Ma\&Mo2017 atmosphere fits}

   The Ma\&Mo2017 atmosphere models are independent of evolutionary models and thus have no a priori reason to be consistent with our bolometric flux measurments. However the best overall fit of a 775~K, intermediate gravity (log $g$=4.5), 13 Jupiter mass object with a radius of 1.07~R$_{\rm Jup}$ is a very good match to the evolutionary model predictions of the observed bolometric flux of \obj for an age of 500~Myr, see Table \ref{Teffage}. The high metallicity solution is very similar to the one obtained with BT-Settl, but with a smaller radius, that is consistent with an older age, closer to 5~Gyr.  As highlighted in section 4.3, both the low gravity and the high gravity Ma\&Mo2017 atmosphere model best fits have associated radius that are too small, in significant tension with the observed absolute flux. When this is considered from the absolute bolometric flux point of view this translates into modelled effective temperature too high to match the evolutionary model predictions.

%


\section{About the nature of CFBDSIR~2149}
  Though the large amount of data we collected clearly identify CFBDSIR~2149 as a peculiar late-T dwarf, it is more difficult to ascertain what kind of peculiar object it is. We therefore discuss in the following subsections  the respective strengths and weaknesses of the four main hypotheses we envision.

\subsection{Is CFBDSIR~2149 a young planetary mass object?}
   The BT-settl model with solar metallicity, very low gravity (log $g$ =3.5), very large radius (1.53 R$_{\rm Jup}$) and a temperature of 700~K is the solar metallicity model which provides the best fit of the spectrum and photometry of CFBDSIR~2149 (Fig. \ref{fullrange}). Qualitatively, the low gravity strongly decreases the collision-induced absorption by H$_2$ in the $K$ band, explaining the red $J-Ks$ colour and the bright $K$-band absolute flux. The low temperature explains the  blue $J-H$ colour and the  strong methane and water absorption band in the NIR spectra. The very large radius is necessary to account for the strong flux emitted by this object that appears much more distant and so intrinsically brighter than what would be expected for a field gravity object. Fig. \ref{colspt} shows the absolute magnitude of CFBDSIR~2149 compared to other field T dwarfs. The \citet{Dupuy.2012} polynomial fit to the absolute magnitude for a field brown dwarf of spectral type T7.5 is fainter by  0.3, 0.3 and 1.0 magnitudes in $J$, $H$ and $K$ bands, respectively. \obj is thus signficantly overluminous only in $K$ band. Even an earlier type T7 field brown dwarf is $\sim$0.5 magnitudes less luminous in $K$ band than CFBDSIR~2149. Our analysis in section \ref{secabsmag} using the bolometric luminosity also shows that a young-age hypothesis and planetary mass range (20 to 500 Myr), results in effective temperatures and radii that are in reasonable agreement with those derived in section \ref{secspectro} from the comparison with atmosphere models. Both BT-Settl models and Ma\&Mo2017 models best fit correspond to young planetary mass objects, with the former favouring very young ages and masses of 2--7\Mjup, and the latter hinting at slightly older ages and heavier masses, up to 500~Myr and 13\Mjup.  We caution that temperatures derived from fitting model atmospheres to normalised spectra and those derived using luminosities and evolutionary model radii do not always quantitatively agree for late-T dwarfs \citep[e.g.,][]{Dupuy.2013}.  Therefore, CFBDSIR~2149 could be somewhat younger or older than we find through this comparison of temperatures, depending on the level of systematic error in the models. \\
   Still, two other independent points raised during our analysis also support a low-gravity in the atmosphere of CFBDSIR~2149. The first is the equivalent width of the K\,{\sc i} doublet at 1.25~$\mu$m that agrees much better with the best fitting low-gravity model than with higher gravity or higher metallicity models. The second is the comparison with the companion GJ~758B whose age (and hence field-gravity log $g$ =5.0--5.5) and slightly super-solar metallicity is known from its main sequence G-type host star. Although the two objects have the same spectral type, the SED of GJ~758B and \obj differ significantly in bands where gravity strongly influences the emergent flux (in $K$ band by collision-induced absorption of H$_2$, and in $Y$ band by the pressure broadened red wing of the 0.77~$\mu$m K\,{\sc i} doublet). These differences in both $Y$ and $K$ bands tend to show that \obj has a lower gravity than GJ~758B, while their very similar colours in $J-H$ confirm that they have similar effective temperatures. The fact that \obj is more luminous than GJ~758B would therefore also be consistent with \obj having a larger radius (and lower gravity) than GJ~758B. Finally, we note that models with both high metallicity and low gravity actually provide the best fit to the data, making it plausible that \obj could be a metallicity-enhanced isolated planetary-mass object.\\
  The main weakness of the young age hypothesis is the lack of any independant age estimation for CFBDSIR~2149, especially because the kinematic data we present in this article clearly shows that it is not a member of any known young association. Though isolated young objects exist in the solar neighbourhood, they are rare.

\subsubsection{Is \obj an isolated analog to 51~Eri~b?}
   The lower end of our inferred age range for \obj (20--100~Myr) is consistent with that of the recently discovered exoplanet 51~Eri~b \citep{Macintosh.2015} that is a member of the $\beta$~Pictoris young moving group \citep[$24\pm3$~Myr;][]{Bell.2015}.  Table \ref{51_Eri} recapitulates the published NIR photometry of  51~Eri~b and compares it with CFBDSIR~2149, which appears to be 0.7--1.0 magnitudes brighter in absolute flux in $J$ and $H$ bands.  According to the ``hot start'' Cond models, such a flux difference at a fixed age of 20~Myr would correspond to masses of 2~\Mjup\ for 51~Eri~b and 3~\Mjup\ for CFBDSIR~2149.  They could even share the same mass if \obj were an unresolved, equal-mass binary, but observational evidence does not currently exist to support or fully disprove such a hypothesis. The $J-H$ colours of 51~Eri~b and CFBDSIR~2149 match within 1$\sigma$, which is indicative that these objects might share a similar SED, but this constraint is not very strong because the errors bars on the 51~Eri~b photometry are very large.  However, new analysis of  $K$-band data of 51~Eri~b from SPHERE suggests it also shares with CFBDSIR~2149 an atypically red $J-K$ colour (Matthias Samland, private communication, Samland et al., submitted.). Since young, L-type exoplanets are usually underluminous in the NIR compared to older objects with a similar effective temperature \citep[e.g.,][]{Skemer.2011}, the similar colours and higher luminosity of \obj could be consistent with it being older than 20 Myr and more massive than 51~Eri~b. However, it is unclear whether such a  trend exists for T spectral types, and the only known young T~dwarf with a parallax and a well established age SDSS~J111010.01+011613.1, a member of AB~Doradus \citep[$149^{+51}_{-19}$~Myr;][]{Bell.2015}, shows no such underluminousity \citep{Gagne.2015c}. Overall, the hypothesis that \obj is a slightly more massive analog to 51~Eri~b is  compatible with  our derived mass and age range for \obj and with the available data for 51~Eri~b, which may be even redder in $J-K$ than \obj given that it is likely lower gravity.

\begin{table}
\caption{Absolute magnitudes and $J-H$ colour of CFBDSIR~2149 and 51~Eri~B  \label{51_Eri}}
\begin{tabular}{|l|c c c|} \hline 
      &  $M(J)$           & $M(H)$           &  $J-H$    \\ \hline \hline
\obj $^1$ &  15.78$\pm$0.04 & 16.14$\pm$0.06  & 0.36$\pm$0.07 \\ \hline 
51~Eri~B $^2$ & 16.75$\pm$0.40 & 16.86$\pm$0.21 & 0.11$\pm$0.45  \\ \hline 
\end{tabular}
\tablefoot{ $^1$The uncertainty on the parallax of \obj corresponds to an additional $\pm$0.2~mag systematic error. $^2$ From \citet{Macintosh.2015} }
\end{table}

\subsection{Is CFBDSIR~2149 a relatively young super-solar metallicity brown dwarf?}
  High-metallicity model atmospheres at moderate gravity provide as good a $\chi ^2$ as the low gravity ones, so it is plausible that \obj could be relatively young with super-solar metallicity. This hypothesis is also consistent with our comparison of \obj to the slightly super-solar metallicity, field gravity (log $g$  = 5.0--5.5) T~dwarf GJ~758B.  Their SEDs disagree but could perhaps be brought into qualitative agreement if \obj has a moderately lower gravity (log $g$ = 4.5--5.0). The main observational evidence that is not compatible with CFBDSIR~2149 being a relatively young super-solar metallicity brown dwarf is its relatively weak K\,{\sc i} doublet at 1.25~$\mu$m that models predict would be much stronger if the object were metal-enriched. However the reliability of atmosphere models in this poorly constrained temperature and metallicity range is not yet established, and direct comparison with the K\,{\sc i} doublet of GJ~758B is not possible because of the lack of resolved spectropic data for this close companion. Independently of these considerations, a significant issue with this hypothesis is that high metallicity objects are very rare, with \cite{Boone.2006} finding that less than 1\% of objects in the solar neighbourhood have [M/H]$>$0.2.

\subsection{Is CFBDSIR~2149 a peculiar unresolved binary?}
   Brown dwarfs with atypical colours and/or spectral features can sometimes be explained by unresolved binarity \citep[e.g.,][]{Bardalez.2014} when the fainter unresolved companion makes a significant contribution at some wavelengths while most of the combined-light spectrum is representative of the brighter primary. In the case of CFBDSIR~2149 the atypical flux excess is in the $K$ band, where the flux is supposed to go down with temperature in the T dwarf range. This implies a higher temperature, higher $K$-band flux companion, but such an object would then be the primary component of the system and therefore dominate the overall SED.  This possibility is excluded by the clear late-T spectral type in the $Y$, $J$ and $H$ bands, as well as with the strong CH$_4$ absorption band at 2.2~$\mu$m. There are other unlikely but still plausible binary scenarios that could account for part of the spectral pecularity of CFBDSIR~2149, such as a partially obscured late-L or early-T companion or even a Y dwarf companion with atypically strong $K$-band flux. The fact that CFBDSIR~2149 is significantly brighter than GJ~758B may favour an hypothesis that \obj is an unresolved equal-mass binary, but then the components would be nearly equal in flux at all wavelengths and thus not explain the atypical shape of their individual SEDs. {Finally we point out that the existence of any lower-mass, cooler companion would be most visible in the thermal infrared, however, as can been seen in Figure \ref{colspt} lower right panel, the IRAC absolute photometry of \obj is exactly what is expected for a sinbgle late T dwarf. This makes the binarity scenario very unplausible. }

\subsection{Is CFBDSIR~2149 an unusually dusty brown dwarf?}

   Dust enhancement has long been used to explain atypically red colours in L and early-T dwarfs, and dust reddenning could account for some of the trends observed in the difference between the best fit and observed spectra (see Fig. \ref{fullrange}).  Without dust, the models used here have a flux excess in the bluer part of the spectrum and a flux deficit in the redder parts compared to CFBDSIR~2149. The work of \citet{Marocco.2014} shows that atypically red L dwarfs can be remarkably well fitted to typical L~dwarf templates of the same spectral type after a simple dereddening by iron-corundrum dust. This hypothesis could explain the very red $J-K$ colour and does provide quite good fit to the overall spectrum of CFBDSIR~2149 (F. Marocco, private communication), but it does not fully account for its blue $J-H$ colour. Indeed, reddening by dust is correlated with wavelength and should therefore also cause a reddening of the $H$ band. Another issue is that it seems difficult to explain how dust could be maintained above the photosphere of such a cold late-T atmosphere without settling. The dust hypothesis also struggles to account for the weak K\,{\sc i} doublet observed in the spectra of CFBDSIR~2149 (table \ref{linewidth}) because enhanced dust abundance would not affect such narrow spectral features despite its potentially strong impact on the overall SED. Finally we note that the best fits with the Ma\&Mo2017 model grid, which include a parameter for the condensation speed of dust always converged toward either the fastest dust condensation speed, leading to very few dust in the photosphere, or to the \textit{no cloud} setting, which explicitely removes all dust in the atmosphere. This is strongly at odds with the hypothesis that CFBDSIR~2149 would be an unusually dusty brown dwarf.

\section{Conclusions}

We have conducted a multi-instrument, multi-wavelength follow-up of CFBDSIR~2149. We have determined its parallax (corresponding to a distance of 54.6$\pm$5.4\,pc), and derived its 6-dimensional position and kinematics. These results show that it is very unlikely that \obj is a member of the AB~Doradus moving group, as claimed by \citet{Delorme.2012a}, therefore removing any strong independent constraint on its age. We also obtained deep NIR spectroscopic observations as well as \textit{Spitzer} photometry in the mid-infrared. Together with the knowledge of its distance, this allowed us to carry out an in-depth spectral analysis of CFBDSIR~2149, notably using absolute and bolometric fluxes, confirming its peculiar nature. Our conclusions are that \obj is most probably either a young ($<$500~Myr) isolated planetary-mass (2--13\,\Mjup) object of late-T spectral type, or an older(2--3~Gyr), metallicity-enhanced, 2--40\,\Mjup, brown dwarf. Our theoretical understanding of cool, low-gravity and/or metallicity-enhanced atmospheres is not yet robust enough to decisively discriminate between these two hypotheses, especially because these physical parameters have very similar effects on the emergent spectra of such atmospheres. However, we point out that there is a distinctive impact on the $J$-band K\,{\sc i} doublet, that does not appear to be significantly affected by low gravity at constant effective temperature, while high metallicity strongly increases its equivalent width. Thus, the K\,{\sc i} doublet could be a crucial tool for discriminating between low-gravity planetary-mass objects and high-metallicity brown dwarfs. In the case of CFBDSIR~2149, the relatively weak K\,{\sc i} doublet favours the hypothesis that it is a young, low-gravity planetary mass object.  Good low to intermediate gravity fits to the data can be obtained with BT-settl models for both solar and [M/H]=0.3 metallictiy and for solar metallicity only with Ma\&Mo2017 models. If \obj is a higher gravity, higher mass brown dwarf then our data are only consistent with models at super-solar metallicity, which is a relatively rare occurence in the solar neighboorhood. Finally, we point out that \obj may be similar in spectral properties, and perhaps also in mass and age, to the recently discovered exoplanet 51~Eri~b \citep{Macintosh.2015}.



\begin{acknowledgements}
 We thank the anonymous referee for his/her valuable comments on our article.
Based on observations obtained with MegaPrime/MegaCam, a joint project
of CFHT and CEA/DAPNIA, at the Canada-France-Hawaii Telescope (CFHT)
which is operated by the National Research Council (NRC) of Canada,
the Institut National des Science de l'Univers of the Centre National
de la Recherche Scientifique (CNRS) of France, and the University of
Hawaii. This work is based in part on data products produced at
TERAPIX and the Canadian Astronomy Data Centre as part of the
Canada-France-Hawaii Telescope Legacy Survey, a collaborative project
of NRC and CNRS.  
''This research has made use of the NASA/ IPAC Infrared Science
Archive, which is operated by the Jet Propulsion Laboratory,
California Institute of Technology, under contract with the National
Aeronautics and Space Administration." 
We acknowledge financial support from ''Programme National de Physique Stellaire" (PNPS) of CNRS/INSU, France. We acknowledge financial support from the French ANR GIPSE, ANR-14-CE33-0018.
\end{acknowledgements}

\bibliographystyle{aa}
\bibliography{biball}

\end{document}